\input harvmac
\noblackbox
\ifx\answ\bigans
\magnification=1200\baselineskip=14pt plus 2pt minus 1pt
\else\baselineskip=16pt 
\fi


\def\tb{type $IIB$\ }
\def\ap{\alpha'}

\def\cf{{\it cf.\ }}
\def\ie{{\it i.e.\ }}

\def\eqq{{\it Eq.\ }}
\def\eqqs{{\it Eqs.\ }}

\def\al{\alpha}

\def\Om{\Omega}
\def\om{\omega}
\def\be{\beta}
\def\Om{\Omega}

\newif\ifnref

\def\doubref#1#2{\refs{{#1},{#2} }}
\def\threeref#1#2#3{\refs{{#1},{#2},{#3} }}

\nreffalse



\def\appC{C}
\def\tilde{\widetilde}

\def\h {{1\over 2}}

\def\ov {\overline}

\def\fc#1#2{{#1 \o #2}}

\def\IZ{ {\bf Z}}
\def\IP{{\bf P}}
\def\IR{ {\bf R}}
\def\hat{\widehat}

\def\br{\hfill\break}

\def\det {{\rm det}}

\def\lf {\left}
\def\ri {\right}
\def\ra {\rightarrow}

\def\im {{\rm Im}}
\def\p {\partial}

\def\Fc {{\cal F}}

\def\Mc {{\cal M}} 
 \def\Tc {{\cal T}}


\lref\GL{T.~W.~Grimm and J.~Louis,
``The effective action of type IIA Calabi-Yau orientifolds,''
arXiv:hep-th/0412277.}
\lref\BKQ{
C.~P.~Burgess, R.~Kallosh and F.~Quevedo,
``de Sitter string vacua from supersymmetric D-terms,''
JHEP {\bf 0310}, 056 (2003)
[arXiv:hep-th/0309187].}
\lref\HL{M.~Haack and J.~Louis,
``M-theory compactified on Calabi-Yau fourfolds with background flux,''
Phys.\ Lett.\ B {\bf 507}, 296 (2001)
[arXiv:hep-th/0103068].}
\lref\JL{H.~Jockers and J.~Louis,
``The effective action of D7-branes in N = 1 Calabi-Yau orientifolds,''
Nucl.\ Phys.\ B {\bf 705}, 167 (2005)
[arXiv:hep-th/0409098].
}
\lref\JLii{H.~Jockers and J.~Louis, to appear.}
\lref\CIUold{
P.G.~Camara, L.E.~Ibanez and A.M.~Uranga,
``Flux-induced SUSY-breaking soft terms,''
Nucl.\ Phys.\ B {\bf 689}, 195 (2004)
[arXiv:hep-th/0311241].
}
\lref\GGJL{
M.~Grana, T.W.~Grimm, H.~Jockers and J.~Louis,
``Soft supersymmetry breaking in Calabi-Yau orientifolds with D-branes and fluxes,''
Nucl.\ Phys.\ B {\bf 690}, 21 (2004)
[arXiv:hep-th/0312232].
}

\lref\CIU{
P.G.~Camara, L.E.~Ibanez and A.M.~Uranga,
``Flux-induced SUSY-breaking soft terms on D7-D3 brane systems,''
arXiv:hep-th/0408036.
}

\lref\KST{
S.~Kachru, M.~B.~Schulz and S.~Trivedi,
``Moduli stabilization from fluxes in a simple IIB orientifold,''
JHEP {\bf 0310}, 007 (2003)
[arXiv:hep-th/0201028].}
\lref\ABFPT{I.~Antoniadis, C.~Bachas, C.~Fabre, H.~Partouche and T.R.~Taylor,
``Aspects of type I - type II - heterotic triality in four dimensions,''
Nucl.\ Phys.\ B {\bf 489}, 160 (1997)
[arXiv:hep-th/9608012].}
\lref\RB{
D.~L\"ust,
``Intersecting brane worlds: A path to the standard model?,''
Class.\ Quant.\ Grav.\  {\bf 21}, S1399 (2004)
[arXiv:hep-th/0401156]; R.~Blumenhagen,
``Recent progress in intersecting D-brane models,''
arXiv:hep-th/0412025.}
\lref\CGP{S.~Cecotti, L.~Girardello and M.~Porrati,
``Two Into One Won't Go,''
Phys.\ Lett.\ B {\bf 145}, 61 (1984).}
\lref\AAS{F.~Ardalan, H.~Arfaei and M.~M.~Sheikh-Jabbari,
``Noncommutative geometry from strings and branes,''
JHEP {\bf 9902}, 016 (1999)
[arXiv:hep-th/9810072].}
\lref\LMRS{D.~L\"ust, P.~Mayr, R.~Richter and S.~Stieberger,
``Scattering of gauge, matter, and moduli fields from intersecting branes,''
Nucl.\ Phys.\ B {\bf 696}, 205 (2004)
[arXiv:hep-th/0404134].
}
\lref\FGP{
S.~Ferrara, L.~Girardello and M.~Porrati,
``Minimal Higgs Branch for the Breaking of Half of the Supersymmetries in N=2
Supergravity,''
Phys.\ Lett.\ B {\bf 366}, 155 (1996)
[arXiv:hep-th/9510074].}
\lref\cupl{G.~Curio, A.~Klemm, D.~L\"ust and S.~Theisen,
``On the vacuum structure of type II string compactifications on  Calabi-Yau
spaces with H-fluxes,''
Nucl.\ Phys.\ B {\bf 609}, 3 (2001)
[arXiv:hep-th/0012213].}
\lref\FKP{S.~Ferrara, C.~Kounnas and M.~Porrati,
``General Dimensional Reduction Of Ten-Dimensional Supergravity And
Superstring,''
Phys.\ Lett.\ B {\bf 181}, 263 (1986).}
\lref\SP{W.~Lerche, P.~Mayr and N.~Warner,
``Holomorphic N = 1 special geometry of open-closed type II strings,''
arXiv:hep-th/0207259;
W.~Lerche and P.~Mayr,
``On N = 1 mirror symmetry for open type II strings,''
arXiv:hep-th/0111113.}
\lref\Beckers{
K.~Becker and M.~Becker,
``M-Theory on Eight-Manifolds,''
Nucl.\ Phys.\ B {\bf 477}, 155 (1996)
[arXiv:hep-th/9605053].}
\lref\Mor{D.R. Morrison, 
``Some remarks on the moduli of K3 surfaces'', Progress in Math., vol. 39, 
Birkh\"auser, 1983, pp. 303-332.}
\lref\MV{D.~R.~Morrison and C.~Vafa,
``Compactifications of F-Theory on Calabi--Yau Threefolds -- I,II,''
Nucl.\ Phys.\ B {\bf 476}, 437 (1996) [arXiv:hep-th/9603161];
Nucl.\ Phys.\ B {\bf 473}, 74 (1996) [arXiv:hep-th/9602114].}
\lref\joe{J.~Polchinski,
``String theory, Vol. 2'', Cambridge University Press, 1998.}
\lref\KW{K.~Wendland,
``Consistency of orbifold conformal field theories on K3,''
Adv.\ Theor.\ Math.\ Phys.\  {\bf 5}, 429 (2002)
[arXiv:hep-th/0010281].}
\lref\Ak{P.~S.~Aspinwall,
``K3 surfaces and string duality,''
arXiv:hep-th/9611137.}
\lref\TV{T.~R.~Taylor and C.~Vafa,
``RR flux on Calabi-Yau and partial supersymmetry breaking,''
Phys.\ Lett.\ B {\bf 474}, 130 (2000)
[arXiv:hep-th/9912152].}
\lref\PMsp{P.~Mayr,
``On supersymmetry breaking in string theory and its realization in brane
worlds,''
Nucl.\ Phys.\ B {\bf 593}, 99 (2001)
[arXiv:hep-th/0003198].}
\lref\GKP{S.~B.~Giddings, S.~Kachru and J.~Polchinski,
``Hierarchies from fluxes in string compactifications,''
Phys.\ Rev.\ D {\bf 66}, 106006 (2002)
[arXiv:hep-th/0105097].}
\lref\Wos{E.~Witten,
``Chern-Simons gauge theory as a string theory,''
Prog.\ Math.\  {\bf 133}, 637 (1995)
[arXiv:hep-th/9207094].}
\lref\AV{M.~Aganagic and C.~Vafa,
``Mirror symmetry, D-branes and counting holomorphic discs,''
arXiv:hep-th/0012041.}
\lref\morr{Morrison on K3}
\lref\vafaf{C.~Vafa,
``Evidence for F-Theory,''
Nucl.\ Phys.\ B {\bf 469}, 403 (1996)
[arXiv:hep-th/9602022].}
\lref\SenVD{
A.~Sen,
``F-theory and Orientifolds,''
Nucl.\ Phys.\ B {\bf 475}, 562 (1996)
[arXiv:hep-th/9605150].
}
\lref\hetsug{
B.~de Wit, V.~Kaplunovsky, J.~Louis and D.~L\"ust,
``Perturbative couplings of vector multiplets in N=2 heterotic string vacua,''
Nucl.\ Phys.\ B {\bf 451}, 53 (1995)
[arXiv:hep-th/9504006];\br
I.~Antoniadis, S.~Ferrara, E.~Gava, K.~S.~Narain and T.~R.~Taylor,
``Perturbative prepotential and monodromies in N=2 heterotic superstring,''
Nucl.\ Phys.\ B {\bf 447}, 35 (1995)
[arXiv:hep-th/9504034].
}
\lref\SVW{
S.~Sethi, C.~Vafa and E.~Witten,
``Constraints on low-dimensional string compactifications,''
Nucl.\ Phys.\ B {\bf 480}, 213 (1996)
[arXiv:hep-th/9606122].}
\lref\GVW{
S.~Gukov, C.~Vafa and E.~Witten,
``CFT's from Calabi-Yau four-folds,''
Nucl.\ Phys.\ B {\bf 584}, 69 (2000)
[Erratum-ibid.\ B {\bf 608}, 477 (2001)]
[arXiv:hep-th/9906070].}
\lref\DRS{
K.~Dasgupta, G.~Rajesh and S.~Sethi,
``M theory, orientifolds and G-flux,''
JHEP {\bf 9908}, 023 (1999)
[arXiv:hep-th/9908088].}
\lref\TT{
P.~K.~Tripathy and S.~P.~Trivedi,
``Compactification with flux on K3 and tori,''
JHEP {\bf 0303}, 028 (2003)
[arXiv:hep-th/0301139].}
\lref\GKTT{
L.~G\"orlich, S.~Kachru, P.~K.~Tripathy and S.~P.~Trivedi,
``Gaugino condensation and nonperturbative superpotentials in flux
compactifications,''
arXiv:hep-th/0407130.}
\lref\ADFL{L.~Andrianopoli, R.~D'Auria, S.~Ferrara and M.~A.~Lledo,
``4-D gauged supergravity analysis of type IIB vacua on $K3 \times T^2/\IZ_2$,''
JHEP {\bf 0303}, 044 (2003)
[arXiv:hep-th/0302174].}
\lref\ADFT{C.~Angelantonj, R.~D'Auria, S.~Ferrara and M.~Trigiante,
``$K3 \times T^2/\IZ_2$ orientifolds with fluxes, open string moduli and critical
points,''
Phys.\ Lett.\ B {\bf 583}, 331 (2004)
[arXiv:hep-th/0312019].}
\lref\thebible{L.~Andrianopoli et al.,
``N = 2 supergravity and N = 2 super Yang-Mills theory on general scalar
manifolds: Symplectic covariance, gaugings and the momentum map,''
J.\ Geom.\ Phys.\  {\bf 23}, 111 (1997)
[arXiv:hep-th/9605032].}
\lref\IIBsugra{
J.~Michelson,
``Compactifications of type IIB strings to four dimensions with  non-trivial
classical potential,''
Nucl.\ Phys.\ B {\bf 495}, 127 (1997)
[arXiv:hep-th/9610151];\br
G.~Dall'Agata,
``Type IIB supergravity compactified on a Calabi-Yau manifold with  H-fluxes,''
JHEP {\bf 0111}, 005 (2001)
[arXiv:hep-th/0107264].}
\lref\LRSi{D.~L\"ust, S.~Reffert and S.~Stieberger,
``Flux-induced soft supersymmetry breaking in chiral type IIb orientifolds
with D3/D7-branes,''
arXiv:hep-th/0406092.
}
\lref\LRSii{D.~L\"ust, S.~Reffert and S.~Stieberger,
``MSSM with soft SUSY breaking terms from D7-branes with fluxes,''
arXiv:hep-th/0410074;\br
A.~Font and L.E.~Ibanez,
``SUSY-breaking soft terms in a MSSM magnetized D7-brane model,''
arXiv:hep-th/0412150.
}
\lref\stieberg{S.~Stieberger,
``(0,2) heterotic gauge couplings and their M-theory origin,''
Nucl.\ Phys.\ B {\bf 541}, 109 (1999)
[arXiv:hep-th/9807124].
}
\lref\AM{I.~Antoniadis and T.~Maillard,
``Moduli stabilization from magnetic fluxes in type I string theory,''
arXiv:hep-th/0412008.
}
\lref\SOFTref{
L.E.~Ibanez and D.~L\"ust,
"Duality anomaly cancellation, minimal string unification and the effective low-energy Lagrangian 
of 4-D strings,''
Nucl.\ Phys.\ B {\bf 382}, 305 (1992)
[arXiv:hep-th/9202046];\br
V.S.~Kaplunovsky and J.~Louis,
"Model independent analysis of soft terms in effective supergravity and in string theory,''
Phys.\ Lett.\ B {\bf 306}, 269 (1993)
[arXiv:hep-th/9303040].
}

\lref\GM{
G.F.~Giudice and A.~Masiero,
``A Natural Solution To The Mu Problem In Supergravity Theories,''
Phys.\ Lett.\ B {\bf 206}, 480 (1988).
}

\lref\KKP{S.~Kachru and A.K.~Kashani-Poor,
``Moduli potentials in type IIA compactifications with RR and NS flux,''
arXiv:hep-th/0411279.
}
\lref\DKPZ{ J.P.~Derendinger, C.~Kounnas, P.M.~Petropoulos and F.~Zwirner,
``Superpotentials in IIA compactifications with general fluxes,''
arXiv:hep-th/0411276.
}

\def\abstract#1{
\vskip .5in\vfil\centerline
{\bf Abstract}\penalty1000
{{\smallskip\ifx\answ\bigans\leftskip 1pc \rightskip 1pc 
\else\leftskip 1pc \rightskip 1pc\fi
\noindent \abstractfont  \baselineskip=12pt
{#1} \smallskip}}
\penalty-1000}
%

\def\hth/#1#2#3#4#5#6#7{{\tt hep-th/#1#2#3#4#5#6#7}}
\def\nup#1({Nucl.\ Phys.\ $\us {B#1}$\ (}
\def\plt#1({Phys.\ Lett.\ $\us  {B#1}$\ (}
\def\cmp#1({Comm.\ Math.\ Phys.\ $\us  {#1}$\ (}
\def\prp#1({Phys.\ Rep.\ $\us  {#1}$\ (}
\def\prl#1({Phys.\ Rev.\ Lett.\ $\us  {#1}$\ (}
\def\prv#1({Phys.\ Rev.\ $\us  {#1}$\ (}
\def\mpl#1({Mod.\ Phys.\ Let.\ $\us  {A#1}$\ (}
\def\atmp#1({Adv.\ Theor.\ Math.\ Phys.\ $\us  {#1}$\ (}
\def\ijmp#1({Int.\ J.\ Mod.\ Phys.\ $\us{A#1}$\ (}
\def\jhep#1({JHEP\ $\us {#1}$\ (}

\def\subsubsec#1{\vskip 0.2cm \goodbreak \noindent {\it #1} \br}

\def\bb#1{{\bar{#1}}}
\def\bx#1{{\bf #1}}
\def\cx#1{{\cal #1}}
\def\tx#1{{\tilde{#1}}}
\def\hx#1{{\hat{#1}}}
\def\rmx#1{{\rm #1}}
\def\us#1{\underline{#1}}
\def\fc#1#2{{#1\over #2}}
\def\frac#1#2{{#1\over #2}}

\def\br{\hfill\break}
\def\noi{\noindent}

\def\al{\alpha}\def\be{\beta}\def\ga{\gamma}\def\om{\omega}
\def\p{\partial}

\def\ofi{$T^2/\ZZ_2 \times$ K3}
\def\mitem{\item{$\cdot$}}
\def\g{\underline{\rm{G}}}
\def\Wh{\breve{W}}
\def\gt{g_{2,0}}
\def\kf{\tx X_V}\def\kb{X_H}
\def\Pic{{\rm Pic}}

\def\xof{X^\sharp}
\def\h{\fc{1}{2}}
\def\Pit{\tilde{\Pi}}\def\Mt{\tilde{M}}
%
\def\La{\Lambda}\def\Om{\Omega}\def\Si{\Sigma}
\def\la{\lambda}
\def\hide#1{}
\def\ZZ{{\bf Z}}\def\IP{{\bf P}}\def\IR{{\bf R}}\def\CC{\bf C}
\def\Im#1{{\rm Im}\, #1}
\def\zb{{\bar{z}}}


%
\vskip-2cm
\Title{\vbox{
\rightline{\vbox{\baselineskip12pt
\hbox{LMU-ASC 1/05}
\hbox{MPP-2005-3}
\hbox{hep-th/0501139}}}\vskip-1cm}}
{F-theory Flux, Destabilization of Orientifolds}
\vskip -1cm
\centerline{\titlefont and Soft Terms on D7--Branes}\vskip 0.9cm
\abstractfont

\vskip 0.8cm
\centerline{D. L\"ust, $\!^{a,b}$\ \ \ 
P. Mayr, $\!^a$\ \ \ 
S. Reffert, $\!^b$ \ and\ \  
S. Stieberger $\!^a$}
\vskip 1.3cm
\centerline{$^a$ Arnold--Sommerfeld--Center for Theortical Physics,}
\centerline{Department f\"ur Physik, 
Ludwig--Maximilians--Universit\"at M\"unchen,}
\centerline{Theresienstra\ss e 37, D-80333 M\"unchen, Germany}

\vskip0.6cm
\centerline{$^b$  Max--Planck--Institut f\"ur Physik, F\"ohringer Ring 6, D-80805 M\"unchen, 
Germany}
\vskip 0.8cm
\abstract{%
We use F-theory to derive a general expression for the flux potential
of type II compactifications with D7/D3 branes, including open string 
moduli and 2-form fluxes on the branes. 
Our main example is F-theory on K3 $\times$ K3 and its
orientifold limit \ofi.
The full scalar potential cannot be derived from the bulk 
superpotential $W=\int \Omega \wedge G_3$ and generically 
destabilizes the orientifold.  Generically all open and closed string moduli 
are fixed, except for a volume factor. 
An alternative formulation of the problem in terms
of the effective supergravity is given and we construct 
an explicit map between the F-theory fluxes and gaugings. We use the
superpotential to compute the effective action for flux compactifications
on orbifolds, including the $\mu$-term and 
soft-breaking terms on the D7-brane world-volume.
 }
\vskip1cm
\Date{\vbox{\hbox{ {January 2005}}
}}
\goodbreak

\parskip=4pt plus 15pt minus 1pt
\baselineskip=14pt 
\leftskip=15pt \rightskip=15pt
\newsec{Introduction}
Type II compactifications with fluxes and D7/D3 brane systems 
play a prominent role in the search for semi-realistic
4-dimensional string vacua.
The effect of 3-form background fluxes
has  been described by a superpotential \GVW\DRS\GKP
\eqn\sp{
W = \int \Omega \wedge G_3,
}
where $\Omega$ is the holomorphic (3,0)-form on the compactification
manifold $Y$ and $G_3=F_3-S H_3$ the quantized background flux.
In the latter expression, $F_3$ ($H_3$) denotes the flux in the RR (NS)
sector and  $S$ is the complex type IIB dilaton.

Formally, the above expression for the superpotential is identical to that for 
closed string compactifications without D-branes 
on a Calabi--Yau 3-fold $Y$. However,
in the latter case, the periods $\int \Omega$ 
of the holomorphic 3-form $\Omega$ compute
the exact scalar potential \TV\PMsp, whereas the above formula for the
orientifold relies on a supergravity approximation.

In the following we improve on previous results by computing the
flux superpotential for certain D7/D3 orientifolds directly 
from F-theory \vafaf. The set of available fluxes in the F-theory is
considerably larger then those described by the bulk potential, and
includes 2-form fluxes on the D7-branes. 
The vacuum structure of the potential obtained 
from F-theory differs in various respects from studies based on eq.\sp,
most notably

\mitem For general fluxes all moduli are stabilized, 
except for a single volume factor. This includes 
the dilaton, K\"ahler and complex structure moduli as well as D7-brane
moduli.

\mitem  The general flux potential destabilizes the orientifold limit
and drives the theory into a regime with varying dilaton, where the
orientifold approximation is not valid and F-theory is needed.

\mitem The supersymmetric flux can not be assigned Hodge type (2,1)
and there is no strict relation between (2,1) fluxes and mass terms for D7-brane moduli. 

The flux compactification of F-theory can be rephrased also in terms
of an effective gauged supergravity, and we give an explicit dictionary
between the fluxes and certain gaugings. An interesting aspect of the
derivation is that the F-theory geometry provides important physical
information beyond the classical effective supergravity, such as the integral
lattice of charged BPS states and the quantization conditions on the gaugings.

We then turn to  a detailed discussion of the case of the 
K3 orbifold $T^4/\IZ_2$, which arises also as the 
$\cx N=2$ supersymmetric sector of $\cx N=1$ orbifolds and plays
a prominent role in the study of intersecting brane worlds. 
In this orbifold limit we study further the gaugings in 
the supergravity approximation and the flux dependent superpotential.
We present explicitly its dependence on the dilaton, 
complex structure moduli and  open string moduli and 
the dictionary between the string theory fluxes and the 
supergravity gaugings. In particular, the superpotential, 
now also depending on the $D7$--brane moduli, gives rise
to a $\mu$--term. 

These results allow to study the $D=4$ effective $\cx N=1$ 
supergravity action 
with spontaneous supersymmetry breaking due to non--vanishing 
auxiliary $F$--term 
components of the moduli fields. In this framework we calculate 
the flux--induced 
soft supersymmetry breaking terms of the effective 
four--dimensional $D7$--brane gauge theory.
In lines of Giudice--Masiero, the $\mu$--term comprises together with 
supersymmetry breaking terms into an effective supersymmetric mass term, 
which gives masses to the non--chiral fermions of the $D7$--brane moduli superfields. 
The latter depends on $(2,1)$--form fluxes only in the case of vanishing
world--volume $2$--form fluxes on the $D7$--branes, but receives a 
$(1,2)$--dependent piece in the presence of fluxes.
On the other hand, the scalar fields of $D7$--brane moduli, describing their positions,
receive both $3$-- and $2$--from flux dependent contributions.

The organization of this note is as follows. In sect.~2 we compute the 
general potential in F-theory on K3 $\times$ K3 for bulk and
brane fluxes and describe its orientifold and orbifold limits. 
In sect.~3 we study the general vacuum structure of the potential 
and discuss some of the new features in a few examples. In sect.~4 we describe the F-theory potentials
in terms of an effective gauged supergravity and
give an explicit dictionary between the 3-form and 2-form fluxes 
in F-theory and  gaugings in supergravity. In sect.~5 we study in 
detail the orbifold $T^4/\IZ_2$ and derive its supergravity description 
in the presence of bulk $3$--form and world volume $2$--form 
fluxes on the $D7$--branes. In sect.~6 we compute the $\mu$-term
as well as the soft breaking masses on the $D7$-brane arising from
the flux potential. 
Some
necessary details on the K3 manifold and its orientifold and orbifold 
limits are discussed in the appendix.

\newsec{Flux potentials in $M$- and $F$-theory}
To compute flux potentials in F-theory, we can build on the 
work on M-theory compactifications on Calabi--Yau 4-folds $X_4$ in
\Beckers\GVW\DRS\HL. If $X_4$ is elliptically fibered, M-theory on $X_4$
is the same as F-theory on $X_4\times S^1$ \MV. The 4-dimensional
F-theory compactification on $X_4$ arises in the limit of infinite
radius of $S^1$. Some aspects of the F-theory limit 
have been discussed already in \GVW\DRS\GKTT.

In the following we will mostly consider the case $X_4=$ K3 $\times$ K3.
This compactification reduces to the type IIB orientifold on \ofi\ 
in a particular limit \SenVD. We will discuss this orientifold limit
later in some detail.
Instead of a product K3 $\times$ K3
one could also consider a non-trivial fibration of K3 over a 2 complex 
dimensional base.  Some of the following
arguments could be generalized to this case.

In the 4-dimensional limit of F-theory on $X_4$, the two K3 factors
play very different roles. The first K3 factor,
on which F-theory is compactified to 8 dimensions has to be
elliptically fibered \vafaf\ and will be denoted by $\kf$. 
The second factor for the compactification from 8 to 4 dimensions will be
denoted by $\kb$. The notation reflects an important distinction of
how the moduli of the two K3 factors enter in the effective 4-dimensional
theory. The latter is a standard effective $\cx N=2$ supergravity
with a number $n_V$ of vector multiplets and a number $n_H$ of 
hyper multiplets. Up to one exception,
the moduli of the elliptic F-theory manifold $\kf$ end up in vector
multiplets, whereas the moduli of $\kb$ end up in the hypers.%
\foot{The effective supergravity
is discussed in some detail in sect.~4.}
The flux potential will reflect this distinction in a natural way.

\subsec{Superpotentials in M-theory}
Upon $S^1$ compactification, F-theory on the elliptically fibered 
4-fold $X_4$ is equivalent
to M-theory with an elliptic fiber of volume $k_E$ inverse proportional
to the radius
of $S^1$. In M-theory, internal flux of the 4-form field strength $G_4$
induces two different superpotentials \GVW\HL:
\eqn\supM{
W=\int_{X_4}\Om\wedge G_4,\hskip50pt
\Wh=\int_{X_4}J \wedge J\wedge G_4,}
where $J$ is the K\"ahler form and
$\Om$ the holomorphic $(4,0)$-form on $X_4$. By definition,
the superpotentials 
$W$ and $\Wh$ depend on complex structure and K\"ahler moduli
of $X_4$, respectively.

Consider an integral 4-form $G_4$ that is the sum of a product 
of integral 2-forms on the two K3 factors 
\def\etat{\tilde{\eta}}
\eqn\defgf{
G_4=\sum \g^{I \La}\,  \eta_I\wedge\etat_\La, \qquad 
\g^{I\La}\in \ZZ.
}
Here $\eta_I$ denotes a basis of $H^2(\kb,\ZZ)$ and similar quantities
with a tilde are used for the same objects on the elliptic K3, $\kf$.
See app. A for some properties of the K3 cohomology 
and notations. The integral matrix $\g^{I\La}$ characterizes the flux 
on the two K3 factors.

On $X_4=$ K3 $\times$ K3, the 4-form $\Om$ and the K\"ahler form $J$ decompose as
\def\omt{\tilde{\om}}
$$
\Om=\om\wedge \omt, \qquad J=j+\tx j,
$$
where $\om$ and $j$ are the holomorphic 2-form and K\"ahler form
on $\kb$ and similarly for the tilded quantities on $\kf$.
In fact, writing \eqn\triplet{\om=\om^1+i \om^2, \qquad j=\om^3~,} 
the three 2-forms $\om^i$ transform as a triplet 
under the trivial $SU(2)$ part of the holonomy of K3.
Choosing a basis of cycles $\ga_I\in H_2(K3,\ZZ)$ 
dual to $\{\eta_I\}$, one has the triplet of period vectors
\eqn\defkt{
\Pi^{I,x}=\int_{\ga_I} \om^x=\int \om^x\wedge \eta^I=
\int \om^x\wedge\eta_J M^{JI}\ .
}
These will be the central objects in the description of  the full moduli 
dependence of the potential in the following. The indices $I$ 
are raised an lowered with the help of the intersection matrix
$M_{IJ}$ defined in (A.1).
We use also $$\Pi^I=\Pi^{I,1}+i\, \Pi^{I,2}$$ to denote the complex combination
corresponding to the holomorphic 2-form. Similar notations will be
used for the ``upper'' K3, $\kf$.

Note that the definition of the superpotential \supM\ seems to 
select a preferred
direction in  $SU(2)$ corresponding to the K\"ahler form $j$.
This defines a fixed complex structure on $\kb$, as well 
as on the moduli space. This will be important
later when comparing with the effective $\cx N=1$ supergravity.

With the above notations, the  superpotentials become

\eqn\supii{
W=\sum_{I\La} \g_{I \La}\, \Pi^I\ \Pit^\La,\hskip50pt
\Wh=\sum_{I\La} \g_{I \La}\, \Pi^{I,3}\ \Pit^{\La,3}.\qquad
}

\noi 
Thus the potential is simply a homogeneous polynomial of degree (1,1)
in the periods of the two K3 factors. One can now evaluate 
the moduli dependence of the potential by computing the period integrals.
An important property of the K3 periods is that they may serve locally 
as good projective coordinates on the moduli space, except for
certain quadratic constraints. 
It follows that the potentials are also 
inhomogeneous polynomials of degree (2,2) in the moduli.
We will describe an appropriate parametrization of the periods momentarily.

Finally, the contribution of $G_4$ to the membrane tadpole \SVW\ is
$$
N_F=\fc{1}{2}\, \int G_4\wedge G_4 = \fc{1}{2}\, \g^I_{\ \La} \g_I^{\ \La}.
$$

\subsec{Superpotentials in F-theory}
Under certain conditions, the potentials \supii\ lift to 
flux potentials of the F-theory compactification to four dimensions.
It is useful to keep in mind two aspects that are new in
this 4-dimensional F-theory limit, which distinguishes the
two K3 factors. As mentioned already, the periods of the
``upper'' K3, $\kf$, become related to vector multiplets,
whereas the periods of the ``lower'' K3, $\kb$, are related
to hyper multiplets,
$$
\Pi^I \longleftrightarrow \rmx{4d\ hypers},\qquad
\Pit^\La \longleftrightarrow \rmx{4d\ vectors}.
$$ 
Secondly, as further discussed in sect.~3, the superpotential
$W$ can be identified with an $\cx N=1$ $F$-term under certain
conditions, while $\Wh$ becomes related to $D$-terms.

The conditions that the M-theory flux compactification has
an 4d F-theory limit are as follows. 
To lift the 3-dimensional M-theory to F-theory in 4 dimensions,
$\kf$ must be elliptically fibered. The dual $E$ of the 
class of the fiber and the dual $B$ of the class of the $\IP^1$ base
of the fibration define two elements of the so-called Picard group
$$
\Pic(\kf) = H^{1,1}(\kf)\cap H^2(X,\ZZ),
$$
with intersections
\eqn\ebi{
\int E\wedge E=2,\qquad
\int E\wedge B=1,\qquad
\int B\wedge B=0.
}
It has been argued in \GVW\DRS, that a selfdual
4-form flux on $X_4$ must be of the form \defgf\ and in
addition  orthogonal to the 2-plane $U$ spanned by $(E,B)$.
Since $\dim H^2(\kf)=22$, the forms $\etat_\La$ appearing 
in the decomposition \defgf\ are restricted to the 20-dimensional
space $\Gamma_V\subset H^2(\kf,\ZZ)$ 
orthogonal to $U$. The dimension 20 of this space is precisely
the number of $U(1)$ gauge fields in the 8-dimensional
compactification of F-theory on $\kf$. One may
naturally think of the flux matrix $\g^{I\La}$ as describing
the 2-form fluxes $\cx F^\La$ of 20 $U(1)$ gauge fields on the ``lower''
K3 $\kb$.

\subsubsec{Fixing the radius of $S^1$}
Note that the radius of $S^1$ in the compactification from 4 to
3 dimensions is a modulus in the theory without flux.
If one searches for 4d flux vacua with a maximal number of moduli
fixed, one would like to fix also the radius of $S^1$ at infinity.
In fact there is another choice of flux in addition to the 
one discussed above, which is appropriate. One can check from
\ebi\ that a flux component
\eqn\fflux{
G^{(4d)}_4 \sim \eta_I \wedge B,
}
with $\eta_I$ a primitive 2-form on $\kb$, is primitive in the 
F-theory limit.\foot{Recall that a primitive 4-form $G_4$ 
fulfills $J\wedge G_4$ and minimizes the potential $\Wh$ \GVW.} 
Switching on a component \fflux\ induces
a term in $\Wh$ that drives the 3d M-theory to 4d F-theory.

\subsec{Moduli dependence of $W$}
As discussed already the, perhaps somewhat abstract looking, 
formulae \supii\ contain already the full moduli dependence
of the four-dimensional potential. We refer again 
to the app. A for a detailed discussion of the
parametrization of the K3 periods. Only for simplicity of exposition
we assume that the ``lower'' K3 is also elliptically fibered.
In a certain parametrization,
the holomorphic periods $\Pi$ and $\Pit$ take the form (A.7)
\eqn\pvcs{
\Pi^I=
\pmatrix{1\cr-U_1U_2+\h W^aW^a \cr U_1\cr U_2\cr W^a},
\hskip 60pt
\Pit^\La=
\pmatrix{1\cr-S U +\h C^a C^a \cr S\cr U \cr C^a},
}
where $a=1,...16$. 

The 18 complex scalars $(U_i,W^a)$ are members of 
4d (hyper) multiplets that 
parametrize the complex structure of the ``lower'' K3, $\kb$. 
For $W^i=0$, the period vector $\Pi^I$ may describe the 
K3 orbifold $T^2\times T^2/\ZZ_2$, with $U_i$ the two 
complex structure parameters of the two $T^2$'s. On the other
hand, the $W^i$ describe the deformations away from the orbifold.

The 18 complex scalars $(S , U ,C^a)$ are components of 
4d vector multiplets. Let us compare this in direction 
to the spectrum of the type II orientifold \ofi, where the scalars in 
the vector multiplets include the dilaton $S '$, the 
complex structure $ U '$ of $T^2$ and 16 D7-brane positions
$C^{\prime a}$. Here we use primes to distinguish the orientifold fields from the 
moduli appearing in the periods $\Pit^\La$. 
One is tempted to identify these two sets of scalars and identify the 
orientifold limit with $C^a=0$ in \pvcs. However this identification is 
non-trivial as illustrated in app. B; the result of the computation
is that the naive identification of the primed and unprimed field 
can be justified, but the integrality properties are obscured in such a 
parametrization. In other words, the ``period vector'' $\Pi^I$, with $C^a=0$
corresponding to the orientifold,  is not defined w.r.t. a basis that 
generates $H_2(\kf,\ZZ)$ over the integers.
Since the integrality properties can nevertheless be recovered from the known 
transformation to the integral basis, we neglect this issue for the moment 
and take it as granted that
$C^a=0$ describes the orientifold limit, and moreover $S $ and $ U $ can 
be identified with the dilaton and complex structure of $T^2$, respectively.

Using the parametrization \pvcs, the general formula for the F-theory 
superpotential \supii\ becomes
\eqn\supiii{\eqalign{
W\ =\ \ &p_1+p_2\, (\fc{1}{2}\, C^aC^a- U  S )+p_3\, S  +p_4\,  U 
+\sum_{a=1}^{16}p_{4+a} C^a\cr
=\ \ &\int_{\kb}\om \wedge \gt\, ,
}}
where $$
\gt=\eta_I \g^I_{\ \La} \Pit^\La\ , \hskip35pt
p_\La=\Pi^I \g_{I\La}\ .
$$
The notation anticipates that $\gt\in H^2(\kb)$ will be of type $(2,0)$ 
at the minimum, 
as will be discussed later. Both forms of $W$ will be useful when studying the
vacuum configurations.

\subsec{Moduli dependence of $\Wh$}
The superpotential for the K\"ahler moduli is \supii
\eqn\Dtermsmo{
\Wh=(\int_{\kb} j\wedge \eta^I)\ \g_{I\La}\ 
(\int_{\kf} \tx j\wedge \etat^\La).
}
In the F-theory limit, the K\"ahler form on $\kf$ is proportional to
the volume form on the base $\IP^1$ 
of the elliptic fibration.
It follows that the integrals on $\kf$ in the above formula are automatically
zero; however the derivatives $\fc{\p}{\p \xi_\La}\Pi^{\Si,3}$ with respect to
the K\"ahler volumes $\xi_\La\to 0$ are not. With an appropriate choice
of coordinates, the content of the variational equations 
orthogonal to the overall volume is 
\eqn\Dterms{
\p_{\xi_\La}\Wh=\int_{\kb} j \wedge \eta^I \, \g_{I\La}=\Pi^{I,3}(T^i)\, \g_{I\La}=0,
}
As will be further discussed below, 
these equations will correspond to $D$-term equations for 
differences of 
$U(1)$ factors from the D7-branes, that restrict the K\"ahler moduli $T^i$
of $\kb$. 
Note that $D$- and $F$-terms cannot be chosen independently,
since both of them are defined by the same flux matrix $\g_{IJ}$.

\subsec{Orientifold limit and bulk moduli}
There is a large amount of literature on flux potentials of type IIB
orientifolds, based on the study of the bulk superpotential eq.\sp.
To relate these results to the present F-theory picture,
it is instructive to study in some detail the orientifold limit.

The bulk fields of the 8-dimensional type II compactification 
on the orientifold $T^2/\ZZ_2$
comprise two complex moduli, the complex type IIB dilaton $S $ and the 
complex structure modulus $ U $ of $T^2$, as well as one real scalar 
$\al$ parametrizing the volume of $T^2$.
In the K3 moduli space, the orientifold limit can be described by the
following Weierstrass form for the elliptic fibration \SenVD
\eqn\ws{
y^2+x^3+x\hx z^4f+\hx z^6g,\qquad f=a\, \prod_{i=1}^4(z-z_i)^2,
\quad g=b\, \prod_{i=1}^4(z-z_i)^3,
}
where $y,\, x,\, \hx z$ are homogeneous coordinates on the fiber and
$z$ is the coordinate on the base $\IP^1$. The K3 described by eq.\ws\
has four $D_4$ singularities at $z=z_i$. These singularities 
reproduce the non-abelian gauge symmetry $SO(8)^4$ from 4 D7-branes on each of the 
4 orientifold planes, for generic $ U $ and $S $. 

Roughly speaking, 
the complex structure $ U $ is related to the cross ratio of the 4 
branch points $z_i$ of the section $\hx z=0$, while the coupling
$S $ may be identified with the modulus of the elliptic 
curve obtained by scaling $(y,x,z)\to (\la^3y,\la^2x,\la z)$ with
$\la\to\infty$. Up to reparametrizations, there are 16 complex parameters in
eq.\ws\ that deform the K3 away from the orientifold limit. 
These correspond to the 16 independent positions of the D7-branes on $\IP^1$.

In a simple approximation near the orientifold limit, 
one can think of $\kf$ as an orbifold $X^\flat=T^2\times T^2/\ZZ_2$. This 
is not entirely correct, 
as $X^\flat$ has the wrong vanishing cohomology and therefore
cannot describe correctly the D-brane states of the orientifold\foot{A 
different orbifold is discussed in app. B.}. As long
as we are only interested in the bulk fields, we can still use $X^\flat$ as 
a model and rewrite 
the 4-form flux $G_4$ as \GVW\DRS
\eqn\fluxd{\eqalign{
G_4 = G_3\, d\bb w + {\rmx{c.c.}},\hskip60pt
G_3 = G_z\, dz +G_\zb\, d\zb.
}}
Here $dw=dx+S  dy$ and $dz=dx'+ U  dy'$ 
are the complex coordinates with $x^{(\prime)},y^{(\prime)}$ 
real coordinates with period one
on $T^2\times T^2$.
The RR and NS components of the type IIB
$G_3$ flux are then identified as $$G_3=F_3-S  H_3.$$
The bulk potential \sp\ thus
leads to the following expression for the superpotential:
\eqn\sptt{
W=\int \Omega \wedge G_3 =  p_1 +p_2 U +p_3S +p_4S  U ,
}
with $\Om=\om\wedge dz$ and \eqn\aie{
p_\La=\int \om\wedge \al_\La, \hskip40pt \al_\La=\eta_I c^I_{\ \La},
\ \  \La=1,...,4.}
Eq. \sptt\  is the superpotential that has been used in the analysis of 
\TT.

On the other hand, the K3 periods on the orbifold reduce to a simple product
of the periods on the two tori
\eqn\ett{
\Pit^\La \ \ \sim \ \ (1,S )\times 
(1, U )=(1,S , U ,S  U ) \ \ \simeq \ \ 
\biggl(SU(1,1)/U(1)\biggr)^2.
}
Thus the bulk superpotential \sptt\ agrees with the
dependence of the F-theory potential \supiii\
on the fluxes $\g^{I\La},$ $\La=1,...,4$ and these F-theory fluxes are 
identified with fluxes of the 3-form bulk fields $H$ and $F$ in the orientifold
limit. 

\subsec{Superpotential from 2-form fluxes on D7-branes}
Compared to the potential \sptt,
 the general superpotential \supiii\ depends on 16 additional
fluxes. As already noted in \DRS, these fluxes should also have a natural interpretation
in the orientifold, as gauge 2-form fluxes 
\eqn\defFa{\cx
F^a=\eta^I\g_{I,a+4},\qquad a=1,...,16,}
on the 16 D7-branes.

Let us substantiate this claim by giving a different derivation
of the superpotential on the 7-brane in the presence of 2-form
flux, similar to the discussion of D5-branes in \AV. 
Consider the B-model of the type II string on the 3-fold $Y$,
with a B-type brane wrapping $Y$. The gauge theory
on the brane is governed by the holomorphic Chern-Simons action \Wos:
$$
W=\int_Y \Omega\wedge Tr \big(A\wedge \bb \p A+\fc{2}{3}A\wedge A\wedge A\big).
$$
The action on the D7-brane, wrapping a 4-dimensional divisor $D\subset Y$,
can be computed by dimensional reduction, keeping in mind that 
the resulting theory on the brane is topologically twisted.
The reduction is more complicated as in the D5 case because of the 
non-trivial dilaton background. However, near the orientifold 
limit \ofi, one has approximately a constant dilaton
on a Calabi--Yau 3-fold. The complex transverse scalar $v$ is a section of
the normal bundle $N(D)\sim \wedge^2 T^*(D)$, where $T^*(D)$ 
denotes the cotangent bundle. In fact $N(K3)$ is trivial, and
$v$ gets identified with one of the deformations $C^a$ in \pvcs\ away from the 
orientifold limit. The dimensional reduction yields
\eqn\wds{
W_{D7^a}=\int_{D7^a}C^a\ \cx F^a \wedge \om,
}
where we used $\Om=\om\wedge dz$. Thus the holomorphic Chern-Simons action
reduced on each of the D7-branes indeed reproduces the 16 extra terms in \supiii.\foot{A
similar result has been obtained from the BI action in \JLii.}

The general situation with varying dilaton
background is more complicated to compute and probably best described
directly in F-theory. However, one can extend the above computation 
to the case of non-trivial normal bundle, with essentially the same 
result, but with $\om$ replaced by a residuum of $\Om$ along $D$.
This is very similar as in the computations of D5-brane 
superpotentials in \AV\SP,
and in fact closely related by the fact that 
a non-zero superpotential \wds\ for a $(0,2)$ flux $F$ 
may be attributed to an induced D5-brane wrapping a non-holomorphic 
2-cycle in $Y$. 

It is important to note however, that the above distinction into
bulk and brane fields (and fluxes) is special to the orientifold limit.
In general,  the 18 complex fields $S , U $ and $C^a$ determine 
locally a point on the coset manifold 
\eqn\modellk{
\fc{SO(2,18)}{SO(2)\times SO(18)}\supset \biggl(SU(1,1)/U(1)\biggr)_S  
\times \biggl(SU(1,1)/U(1)\biggr)_ U ,
}
which describes the complex structure of the elliptically fibered F-theory K3, $\kf$ \vafaf.
The parametrization \pvcs\ displays explicitly the embedding of the coset of
bulk fields \ett\ at the orientifold point $C^a=0$. 
However, as will be discussed in sect.~3,  when we study the
vacuum equations, the 2-form fluxes on the D7-branes 
destabilize the orientifold and drive the branes to generic positions.
For general values of the fields, 
the distinction into bulk and D7-brane moduli is no meaningful notion
because of the strong back-reaction of the geometry and the dilaton. 
A similar comment applies to the distinction into 4 bulk fluxes and 16 brane fluxes.

\subsec{Orbifold limit of $\kb$}
On the other hand, one can further simplify the orientifold limit
by considering the special case, where also the ``lower'' K3, $\kb$, 
is described by an orbifold. This makes contact to the 
large literature on intersecting branes on toroidal orbifolds, see 
the reviews \RB\ and references therein.

The untwisted moduli space of $Z_N$ orbifolds has been well studied and is 
described by certain cosets described in \FKP. 
K3 orbifolds are included in this
classification as the $\cx N=2$ supersymmetric orbifold sectors with fixed tori.
We will discuss the orbifold limit in detail in sects. 5 and 6, 
where we compute
the effective action including the soft-breaking terms.  Here we restrict
to display the content of the formula \supiii\ for 
the simple situation, where $\kb$ is the orbifold $T^2\times T^2/\ZZ_2$.
As discussed already, the two complex structure moduli $U_i$ 
are described by setting
$W^a=0$ in \pvcs. Plugging $\Pi$ and $\Pit$ into \supiii\ one gets\foot{We are again oversimplifying here, similar as in the case of the orientifold
limit. Although one may describe the $\ZZ_2$ orbifold by a ``period vector''
as in \pvcs, the latter is not defined on a basis that 
generates $H^2(\kb,\ZZ)$ of the orbifold 
over the integers. As a consequence, the entries of the
matrix $\g$ appearing in the following equation are not integers but 
fractional linear combinations of integers. The integrality properties can 
be reconstructed from the explicit transformation from the integral basis
to the orbifold basis, see app. B for more details.}
\eqn\spof{\eqalign{
W_{(T^2\times T^2/\ZZ_2)}=\hskip12pt
1&\, (\g_{11}+\g_{12}(\h\, C^aC^a- U  S )+\g_{13} S +\g_{14} U +\g_{1(a+4)}C^a)-\cr
U_1U_2&\, (\g_{21}+\g_{22}(\h\, C^aC^a- U  S )+\g_{23} S +\g_{24} U +\g_{2(a+4)}C^a)+\cr
U_1&\, (\g_{31}+\g_{32}(\h\, C^aC^a- U  S )+\g_{33} S +\g_{34} U +\g_{3(a+4)}C^a)+\cr
U_2&\, (\g_{41}+\g_{42}(\h\, C^aC^a- U  S )+\g_{43} S +\g_{44} U +\g_{4(a+4)}C^a).
}}

\noi
At $C^a=0$, this is precisely the superpotential that has appeared in the
orbifold literature on intersecting branes \LRSi. Equation \spof\ 
generalizes their bulk superpotential to include also the moduli $C^a$
of the D7-branes. As discussed in \defFa, the fluxes $a>4$ with linear
coefficients in the $C^a$ describe the a special
subset of the 2-form fluxes $\cx F^a$ 
on the D7-branes, with $I=1,...,4$. However we will
argue below that these fluxes destabilize the orientifold and 
can not be incorporated consistently within the orientifold/orbifold model.

\def\f{\underline{\rm{F}}}
\def\HH{\cx G_{flux}}
\def\FF{\cx F_{flux}}
\newsec{Minimization of the effective 4d potential}

\subsec{F-theory potentials in four dimensions}
Before we turn to a study of the 4d vacua with fluxes, it is instructive
to understand some general properties of the two superpotentials
\eqn\onceagain{\eqalign{
W&=\int_{\kb} \om \wedge \eta^I \ \g_{I\La}\  \int_{\kf} \omt \wedge \etat^\La,
\hskip 20pt \Wh= \int_{\kb} j \wedge \eta^I \ \g_{I\La}\  
\int_{\kf} \tx j\wedge \etat^\La.
}}
Since the potential $W$ is complex, whereas $\hx W$ is a real function,
one would like to identify tentatively $W$ with a 4d $F$-term
and $\hx W$ with a $D$-term. However the definition of the potentials
in \onceagain\ is based on the distinction between K\"ahler and 
complex structure moduli of K3, which is an ambigous concept because of the
$SU(2)$ symmetry that rotates the three 2-forms
\triplet\ into each other. 

In the four-dimensional F-theory limit this ambiguity is resolved by the 
zero elliptic fiber limit of the ``upper'' K3, $\kf$. The holomorphic elliptic fibration
of $\kf$ selects unambigously a complex stucture $\omt$ and a K\"ahler 
structure $\tx j$. Moreover all K\"ahler classes of $\kf$ tend to zero 
in the F-theory limit of zero size fiber, and thus the {\it value} of $\Wh$ is
strictly zero for the fluxes orthogonal to the classes $E$ and $B$ of the
fibration in eq.\ebi. 
The 4d scalar potential therefore gets contributions
only from the variations $d\Wh$ as in \Dterms, which is compatible with the interpretation
of as a $\Pi^{I,3}\g_{I\La}$ as a $D$-term in the $\La$-th $U(1)$ factor.
More precisely, a 2-flux proportional to a 2-form $\etat_\La$ 
couples to the difference of two $U(1)$ factors associated with two D7-branes.

The small fiber limit does not resolve a similar ambiguity in the definition of
K\"ahler and complex structure deformations on the ``lower'' K3 $\kb$. This
reflects the fact that these fields are not decoupled in the potentials and
moreover each flux contributes always to both, $F$- and $D$-terms. 
It is also clear that one can not distinguish $F$- and $D$-terms 
by the Hodge decomposition on $H^2(\kb)$. 

In fact, in the compactification without flux, the $SU(2)$ that
rotates the complex structure on $\kb$ is an $R$-symmetry and the 
only effect that breaks this symmetry is the flux potential itself.
We are thus led to define the 
Hodge structure on $\kb$ by the projection to $H^2(\kb)$ of the Hodge structure on the 4-form cohomology $H^4(X_4)\supset G_4$ at the minimum. 
Since the Hodge type of $G_4$ is $(2,2)$
at the minimum \Beckers, the $(2,0)$-form on $\kb$ can be defined as the projection $\mu$
of the component 
$$
\hx G_4=\mu\wedge\bb\omt,
$$
onto $H^2(\kb)$. With this definition, we can associate the complex potential $W$ with
a $F$-term which may give an obstruction to the first order deformations
corresponding to elements in 
\eqn\defi{
H^{2,0}(\kb)\otimes H^{1,1}(\kf)\oplus 
H^{1,1}(\kb)\otimes H^{2,0}(\kf)\subset H^{3,1}(X_4)\, .}
The real $D$-term potential $\hx W$ may obstruct, at first order, 
only the deformations in 
\eqn\defii{
H^{1,1}(\kb)\otimes H^{1,1}(\kf)\subset H^{2,2}(X_4).} 
Note that deformations of 
both type of potentials involve 2-forms of type  $H^{1,1}(\kb)$,
reflecting the fact that the complex structure and K\"ahler 
moduli from $\kb$ are not decoupled.

\subsubsec{Deformations that preserve the complex structure}
In a concrete physical application, it is sometimes important to 
classify the fluxes that preserve a certain complex structure. E.g.,
the orientifold models correspond to a special complex structure
where the D7-branes sit on top of orientifold planes. This is only 
a consistent restriction if the flux potential does not destabilize  
the orientifold limit. The same problem appears in 
orbifold models of K3 that describe only a few of the 
possible directions of the potential.

The above equations can also be interpreted as saying,
that at first order, complex stucture moduli get a potential 
from a flux component of Hodge type \defi, while K\"ahler
moduli get a potential from a component of type \defii. In fact, at second order 
in the deformation, this distinction is lost and each flux component
contributes to both potentials, or $F$- and $D$-terms. However 
if we switch on a ``new'', extra 
flux component $$G_4^*\in H^{1,1}(\kb)\times H^{1,1}(\kf)$$ 
of type \defii\ in addition to $G_4$, 
the contribution  to the $F$-potential will be of the form
\eqn\wextra{
\delta W(G_4^*)\sim \phi^2\ ,
}
where $\phi$ denotes collectively the 
complex structure moduli with value $\phi=0$ at
the original flux vacuum. Thus an extra flux of pure type $\defii$ will preserve
the complex structure at the minimum of the original potential $W$. 
The above situation may be visualized in terms of the integral flux matrix $\g$ as 
\eqn\fluxmatrix{
\g_{I\La} = \pmatrix{
\vbox{\offinterlineskip\tabskip=0pt\halign{\strut
\hfil~$#$~\hfil
&\hfil~$#$~\hfil\vrule&\hfil~$#$~\hfil$
\phantom{\pmatrix{1\cr1}}$\hskip-14pt\cr
\hx \g^{+\,+}_{\ 3\times 2}&
\hx \g^{+\, -}_{\ 3\times n'}&0\cr
\hx \g^{-\, +}_{\ n\times 2}&
\hx \g^{-\, -}_{\ n\times n'}&0\cr
\noalign{\hrule}
0&0& \f^{-\, -}_{\ m\times m'}\cr
}}}
\ ,}
where we ordered the orthonormal 
basis $\{\eta_I\}$ for $H^2(\kb)$ such that the 
elements with positive self-intersection are at the first positions and similarly
for the basis $\{\etat_\La\}$ 
for $H^2(\kf)$. The vectors defined by the left upper corner 
$\hx \g$ span a flux space 
$\HH\subset H^4(X_4,\IR)$ which maps into subspaces 
of signature $(3,n)$ and $(2,n')$ under the projection to  
$H^2(\kb)$ and $H^2(\kf)$, respectively. 
Similarly the right lower corner $\f$ can be associated with an
orthogonal subspace $\FF$ which maps to subspaces of $H^2(X_i)$ of purely negative
signature under the projection to the K3 cohomology.

In the above notation, 
an extra  flux $G_4^*$ that preserves the complex structure at the minimum of the flux
potential defined by $\hx \g$, must be of the type $\f$. Note that the 
extra flux $G_4^*$ fulfills the usual quantization conditions and projects to
integral $(1,1)$ forms on both K3 factors. A necessary condition for the
existence of this type of flux is therefore, that Pic($\kb$) is non-empty 
and that dim(Pic($\kf$))$>2$. 

Although $G_4^*$ does not change the minimum for the complex structure moduli,
the $D$-term condition $G_4^*\wedge j=0$ may impose new conditions on the 
K\"ahler moduli:
\eqn\whextra{
\delta d\Wh(G_4^*) = \Pi^{I,3}\f_{I\La}=0.
}
Let us discuss briefly two important examples of $D$-term conditions.
If Pic($\kb$) is empty, there are still non-trivial $D$-term conditions
from the flux of type $\hx \g$ in \fluxmatrix. A universal contribution
may arise from a flux aligned with $j$ on $\kb$; however, without 
further contributions to the potential, it drives the theory to the boundary of the moduli 
space.
A more interesting case with a minimum in the interior of the moduli 
space is the following one. 
Let $e_i\in H^2(\kb)$ be the duals of exceptional divisors
in the resolution of an orbifold singularity of $K3$. 
At the singular point, $\om \wedge e_i=0$ and the K\"ahler form 
can grow components in the directions $e_i$. There is a
moduli space of K\"ahler structures with K\"ahler forms 
$$
j'(T^i)=j+\sum_i T^i\ e_i, \qquad e_i\in \Pic(\kb),\ T^i\in \IR,
$$
and K\"ahler moduli $T^i=\Pi^{i,3}=\int_{e_i^*}j'$. A non-zero flux with components 
in the directions of $e_i$ induces a $D$-term potential 
$d\Wh$ which imposes again a linear relation 
\eqn\dex{
\sum T^i\g_{i\La}=0,
}
on the volumes of the $e_i$. Note that this function is 
an arbitrary linear combination of the volumes of the 
exceptional divisors, and the zero does not mean that
an individual exceptional divisor shrinks at the minimum;
therefore the generic minimum is not at an geometric orbifold point
in the moduli.

\subsec{General structure of minima}
We discuss now in more detail the vacuum conditions imposed by 
the potential arising from $W$. 
The equations
\eqn\vari{
D_{n} W =(\p_n +\p_nK)\ W=0,
}
describe minima at a vanishing value of the potential.
Here $n$ runs over all complex hyper and vector moduli\foot{From F-theory
perspective it is natural to include the dilaton amongst the
complex structure moduli and we do so in the following.} 
in the periods $\Pi$ and $\Pit$, respectively, and 
$K$ is the K\"ahler potential. The number of equations
is equal to the number of moduli and thus all moduli appearing
in $W$ are fixed in a generic solution.

It is straightforward to compute the content of \vari\ for the
potential \supii. The result is that if the solution is in the interior 
of the moduli space, the 4-form flux $G_4$ decomposes into 
pieces that are sums of forms of definite Hodge type
\eqn\fluxdec{
G_4 = G_0+G_1+G_2,\hskip30pt \cases{
G_0 
\ \in H^{2,0}_B\wedge H^{2,0}_F+\rmx{c.c.}&\cr
G_1 
\ \in H^{2,0}_B\wedge H^{0,2}_F+\rmx{c.c.}&\cr
G_2 \in H^{1,1}_{B}\wedge H^{1,1}_{F}&\cr
}}
whereas the components in \defi\ are set to zero by the constraints \vari. 
The potential thus reproduces part of the general vacuum conditions of 
\Beckers\ and was designed to do so in \GVW.

The general configuration describes a non-supersymmetric vacuum with vanishing
potential. There are also two supersymmetric  branches of the vacuum space, characterized by
the vanishing of components in the decomposition \fluxdec. As there are no more
moduli to tune, a vanishing of these components can only be achieved by treating the 
entries of the flux matrix $\g_{I\La}$ as further parameters;
however such a tuning may be in conflict with the flux quantization.  
Neglecting this problem for the moment one
finds the following three different branches:
$$\eqalign{
&\cx N=0:\hskip40pt \rmx{generic}\ ,\cr
&\cx N=1:\hskip40pt G_0=0,\cr
&\cx N=2:\hskip40pt G_0=G_1=0.\cr
}
$$
Each constraint $G_a=0$, $a=0,1$ gives 
two real conditions on the real matrix $\g_{I\La}$.
The $\cx N=1$ supersymmetric branch is characterized by the vanishing of $G_0$,
where $W=0$ in addition to \vari.

On the other hand, for $G_1=G_0=0$, the flux $G_4$ 
is of type $(1,1)$ on both K3's. If it is primitive as well, 
it is orthogonal to all 2-forms $\om^i$ in eq.\triplet. 
Thus it is a primitive $(2,2)$ flux w.r.t. to 
the $\IP^1\times \IP^1$ of complex structures of the M-theory compactification 
without flux, which is a sufficient criterion for extended supersymmetry \DRS.
In the decomposition of the flux matrix \fluxmatrix, this corresponds to the case
$\hx \g=0$, $ \f \neq0$.

\subsec{Vacuum solutions}
To determine the $\cx N=1$ supersymmetric vacua, we consider the potential 
\supiii\ in the form,
$$
W=\int_{\kb} \om\wedge g_{2,0},\qquad g_{2,0}=\eta_I\, \g^I_{\ \La}\, \Pit^\La \, .
$$
Since $D_i \om$ is a form of pure type $(1,1)$, 
the equations $W=D_iW=0$ imply that $g_{2,0}$ is of type $(2,0)$. Comparing this
with the expansion $\om=\Pi^I\eta_I$, we find
\eqn\vaccondi{
\Pi^I=c\ \g^I_{\ \La} \Pit^\La, \qquad c\in \CC
}
at the vacuum. Inserting this into the quadratic constraint $\om\wedge\om=0$, one finds
the following equation, which is equivalent to $W=0$:
\eqn\wen{
\Pit^\Si \g^I_{\ \Si}\g_{I\La}\Pit^\La=0=\Pit^\La \Mt_{\Si\La}\Pit^\La.
}
On the r.h.s. we have written the quadratic constraint on the periods of $\kf$ to 
illustrate the two basic ways to solve this equation. A 
special solution is
\eqn\symm{
\g_{\La\Si}=\Mt_{\La\Si}\ \  \longleftrightarrow\ \  \Pi^I\sim \Pit^\La
}
This is a symmetric solution, where the two K3 factors are at the same 
moduli\foot{In particular, $\kb$ has an elliptic fibration and the
non-zero period vector has dimension 20, as for $\kf$.}.
Note that this solution to \wen\ 
has a moduli space corresponding to the complex structure of 
one elliptic K3.

The above branch of solutions 
arises from the quadratic constraint satisfied by the periods
for all values of the moduli. The other possibility to satisfy \wen\ is if $\Pit^\La$
is a null vector of $\g$ at special values of the moduli. The generic solution is
a combination of both solutions, that is the contribution of 
non-zero vectors must be proportional to the quadratic constraint.

To determine all moduli, one may use the Hodge decomposition \fluxdec,
explicitly:
\eqn\hdec{\eqalign{
g_{2,0}&=\eta_I\, \g^I_{\ \La}\, \Pit^\La\ ,\cr
g^a_{1,1}&=\eta_I\, \g^I_{\ \La}\, D_a\Pit^\La\ ,\cr
g_{0,2}&=\overline{g_{2,0}}\ ,\cr
}}
where $D_a=\p_a+K_a$ are the covariant derivatives w.r.t. the complex moduli
of $\kf$. The values of the moduli at the vacuum are then determined by 
eq.\wen\ and the equations
\eqn\wenii{
g_{2,0}\wedge g^a_{1,1}=g_{2,0}\wedge \overline{g^a_{1,1}}=0.
}
For the supersymmetric vacua satisfying \wen, these equations simplify 
considerably as one may replace the covariant derivatives by ordinary ones
and moreover the ordinary derivatives are linear in the period vector
$$
\p_a\Pit^I = (X_a)^I_{\ J}\Pit^J,
$$
for a constant matrix $X_a$.
The equations \wen,\wenii\ represents a system of holomorphic 
polynomials of degree $\leq 4$ that is straightforward to solve. We will
discuss some examples below.

\subsubsec{$\cx N=2$ vacua}
The $\cx N=2$ vacua correspond to $G_1=G_0=0$, or 
\eqn\net{
g_{2,0}=\eta_I\, \g^I_{\ \La}\, \Pit^\La=0,}
which is equivalent to $\om\wedge G_4=\omt\wedge G_4=0$.
This is possible only if $\g$ is a sum of products of anti-selfdual
forms on the two K3 factors.
The space of $\cx N=2$ vacua has various branches labeled
by $r=\rmx{rank}\ \g\leq 18$. The residual geometric 
moduli space  on the branch labelled
by the integer $r$ is generically
\eqn\netvac{
\IR_+\times
\fc{SO(3,19-r)}{SO(3)\times SO(19-r)}
\times \fc{SO(2,18-r)}{SO(2)\times SO(18-r)}\times \IR_+.
}
Note that the fact that $\om \perp G_4$ does not necessarily mean
that the $\cx N=2$ solutions are orbifolds,  
since the classes appearing in $G_4$ may have non-zero K\"ahler volumes.
In fact the generic $D$-term potential \Dterms\ has a minimum that 
fixes the K\"ahler volumes away from 
the orbifold.

\subsubsec{Non-supersymmetric vacua}
Non-supersymmetric vacua are described by solutions to eq.\wenii\ that do
not solve eq.\wen. These vacuum equations are no longer holomorphic
polynomials, but can still be solved for specific examples.

\subsec{Examples}
We will now study some features of solutions to the general potential 
\supiii\ in some examples. As discussed already, the equations 
$W=D_i W=0$, where $i$ is a modulus of $\kb$, are solved by 
$\gt \in H^{2,0}(\kb).$ Differentiating the 
first line of \supiii\ w.r.t. the moduli of $\kf$ one finds,
in terms of $p_\La=\Pi^I\g_{I\La}$,
\eqn\vacsol{
\vbox{\offinterlineskip\tabskip=0pt\halign{\strut
$#$~,\hfil\hskip40pt&\hfil~$#$~\cr
p_3-p_2\,  U =0&p_{4+a}+p_2\, C_a=0\, ,\cr
p_4-p_2\,  S =0&p_1+p_2\, ( U  S -\fc{1}{2}\, C^aC^a)=0\ .\cr
}}}
The two branches of solutions to the above equations are
\eqn\vacsolii{
\vbox{\offinterlineskip\tabskip=0pt\halign{\strut
$#$~\hfil\hskip20pt&$#$~\hfil\hskip40pt&$\displaystyle#$\hfil\cr
\underline{\cx N=2:}&p_\La=0&\gt=0~.\cr\cr
\underline{\cx N=1:}&p_2\neq 0& U  = \fc{p_3}{p_2},\qquad  S  = \fc{p_4}{p_2},\qquad C^a = -\fc{p^{4+a}}{p_2}\, .\cr
}}}
The above formula displays already two important properties of the generic 
solution on the $\cx N=1$ supersymmetric branch: 

{$i)$} for generic flux, all $D7$ moduli $C^a$ are fixed.\foot{Near the orientifold limit, 
this follows from the local analysis presented in \GKTT.} 

$ii)$ For generic flux, the $D7$ branes are stabilized at 
generic positions $C^a\neq 0$
away from the orientifold. Thus the orientifold is not
a 'preferred' point in the space of vacua and the generic flux destabilizes
the orientifold geometry. 

\subsubsec{Orientifold/orbifold compactification}
As the first concrete example consider the orientifold. Shifting coordinates
as in (B.2) and after a basis transformation, one obtains the 
``period vector'' $\Pit^\La$ in eq.(B.3) for the orientifold, with the
orientifold point described in the shifted variables as $C^a=0$.
Similarly one may use a transformed ``period vector'' $\Pi^I$ for the
``lower'' K3, $\kb$, such that the point $W^i=0$ corresponds to a point
in the moduli of the K3 orbifold $T^4/\ZZ_2$. Although the
following discussion does not rely on assuming an orbifold
limit for $\kb$, it it is instructive to do so to compare with
the explicit discussion in sect.~5.

In the orientifold/orbifold  basis for the period vectors, 
switch on a flux $\g$ corresponding to
a plane $\HH$ with signature $(2,2)$ defined by $\g_{I\La}=0$ for 
$I>5,\La>4$. 
The potential is given by \supiii, with all $p_\La,$
$\La>4$ zero. The minimization of the potential is then as in \TT, except for
two differences. First recall that the entries of
$\g$ in this basis are not integers, but fractional linear combinations
of integers that can be determined by the transformation to the
integral basis (A.7). Secondly the potential has an additional quadratic piece 
in the $D7$ brane moduli 
$$
\delta W = \fc{1}{2}\, p_2 \sum_a (C^a)^2
$$
which sets $C^a=0$ for all $a$ if $p_2\neq 0$. Recall that $p_2=0$
corresponds to the $\cx N=2$ supersymmetric branch, so the $D7$--branes
are fixed to the orientifold plane by this choice of bulk flux
on the $\cx N=1$ supersymmetric branch. Similarly, the minimum 
for the moduli of $\kb$ is at the orientifold point $W^i=0$.

As discussed before, a generic flux will destabilize the orientifold.
However, as discussed in sect.~3.1, 
one can also add another flux to this compactification, 
without destabilizing the orientifold. 
Consider an additional flux $\f$ corresponding
to another plane $\FF$ such that $\HH$ and $\FF$ are orthogonal,
that is $\f_{I\La}=0$ for $I\leq 5,\La \leq 4$.\foot{As explained earlier, this
requires that Pic($\kb$) is non-empty.}
This flux corresponds to a 2-form flux $$\cx F^a=\f_{ia}\eta^i$$ on the $D7$ branes. 
In the above basis, there is an extra superpotential of the type \wextra,
$$
W' = \sum\, W^i\, C^a\, \f_{ia},
$$
whose value and derivatives are zero at the minimum of the original potential
$W$. Thus a 2-form flux $\cx F^a$ of the type $\f$ induces a superpotential that 
does not destabilize the orientifold/orbifold vacuum.

In addition to the superpotential $W'$, 
the same flux $\cx F^a$ induces an extra  
$D$-term potential \whextra. By definition, 
this $D$-term restricts K\"ahler moduli on $\kb$ 
\eqn\sdtof{
d\Wh'=T^i\ \f_{ia}=0 \ ,\qquad T^i=\Pi^{3,i}\in \IR.
}

\subsubsec{Some more examples with 2-form flux}
In the following we illustrate some of the features of the general
potential at the hand of some simple examples. The fluxes in these
examples are given in the orthonormal basis for both $K3$ factors,
corresponding to the period vectors in the form (A.7).

\mitem{$\underline{G_4=\eta^i\wedge \eta^a,\ i,a>4}$}\br
This choice for $G_4$ corresponds to a pure 2-form flux $\cx F^a=\eta^i$
on the $a$-th brane with rank$(\g)=1$ (and is the special case
of the above example, with $\g\to0,\f\to\g$). The superpotential is
$W=W^i\cdot C^a$, with a $\cx N=2$ supersymmetric minimum 
at $W^i=C^a=0$. The Hodge type of $\cx F^a$ at the minimum is $(1,1)$.
The same flux induces  both, the $F$-term potential $W$ as well as a $D$-term
potential with minimum $T^i=0$. This is the simplest example to 
illustrate that all components of the 2-form flux on the brane 
enter both potentials.
\vskip10pt

\mitem{$\underline{G_4=\eta^i\wedge \eta^4,\ i>4}$}\br
This is again a rank$(\g)=1$ example, which however restricts the bulk fields. 
The superpotential is $W=W^i\cdot ( S - U )$ with an 
$\cx N=2$ supersymmetric minimum at $W^i=0= S - U $. 
There is also a $D$-term that requires $j\perp \eta^i$.
Note that this is a supersymmetric flux vacuum with bulk fluxes 
where nevertheless all the $D7$-brane moduli are massless.
This illustrates that supersymmetric bulk fluxes of type $(2,1)$
need not give 
masses for the $D7$ moduli, differently then in previous examples 
presented in \GKTT\CIU. 
\vskip10pt

\mitem{$\underline{G_4=\eta^I\, M'_{I\La} \etat^\La}$}\br
Here $M'_{I\La}=M_{\Si\La}$ for $I=\Si<19$ and zero for $I=19$.
This is a symmetric solution, where both K3 factors are elliptically
fibered and at the same complex structure moduli. The vacuum 
has $\cx N=1$ supersymmetry and a large complex structure moduli space,
corresponding to one elliptically fibered K3.
\vskip10pt

\subsubsec{Change of hodge type of the supersymmetric flux}
In the case of only a bulk potential \sp\ with a supersymmetric
minimum, the Hodge type of the 3-form flux $G_3$ is $(2,1)$.
This is also reflected in the computations of mass terms on 
the brane \doubref\CIU\LRSii.

In the generic configuration with general fluxes and 
$D7$-branes at arbitrary positions, there is no obvious notion of
what one calls a flux of Hodge type $(2,1)$. However one can ask the 
simpler question, whether at least under a small deformation of the 
$D7$ branes away from the orientifold, the supersymmetric bulk flux 
remains of type $(2,1)$. In fact there is no reason to expect this and 
indeed the following  short computation
shows that the Hodge type changes already at first order in the deformation.
This should also be relevant for computations of soft terms 
on the world volume of D-branes away from the orientifold plane.

The Hodge decomposition for the bulk flux is, in the notation of sect.~2.5, 
$$
G_4 = G_3 d\bb w+G_3^* dw = 
(G_z dz +G_\zb d\zb)\, d\bb w +(G^*_z d\zb +G^*_\zb dz)\, dw.
$$
At the supersymmetric minimum, $G_3$ is of type $(2,1)$ and thus 
$G_z$ is of type $(1,1)$ and  $G_\zb$ of type $(2,0)$.
To include the 2-form fluxes, we first write the Hodge decomposition 
in terms of the $K3$ periods as
$$
\int G_4 \wedge \tx\om = G_\zb \Pi +G_z \Pi_ U +\rmx{c.c.}.
$$
where $\Pi_a=D_a \Pi$ projects to a form of  pure type $(1,1)$ 
on the ``lower'' K3 $\kb$.  To determine the
effect of a small perturbation it suffices to consider the case with 
a single 2-form flux. The Hodge decomposition of the 4-form flux
becomes
$$
\int G_4 \wedge \tx\om = G'_\zb \Pi' +G'_z \Pi'_ U +\cx F^a\, \Pi'_{C^a}\ +\rmx{c.c.}\ ,
$$
where the prime indicates the periods at $v\neq 0$ and $\cx F^a$ is identified
with a non-zero 2-form flux on the brane that induces the deformation. 
In the vacuum, $G'_\zb$ is of type $(2,0)$ while $G'_z$ and $\cx F^a$ are 
of type $(1,1)$. 

To compute the change of Hodge type linearly in the deformation
$C^a$, one may rewrite $G_4$ in terms of the original Hodge decomposition;
at the required order, it is consistent to keep the complex structure of $T^2$ 
constant. The coefficient of the 2-form 
$d\zb d\bb w$ is 
$$
G'_\zb+\fc{1}{4 S _2 U _2}\bb C^a\ (\cx F^a+\bb {\cx F}^a).
$$
Thus the supersymmetric bulk flux is no longer of
pure type $(2,1)$ but has a piece of 
Hodge type $(1,2)$ at first order in the deformation.

\newsec{Embedding in supergravity}
Below we rephrase the flux potential \supiii\ in the language of 
the effective 4d supergravity and give a direct relation between
the flux parameters in \defgf\ and certain gaugings in supergravity.
Aspects of the effective supergravity for the orientifold
limit have been discussed in \DRS\TT\ 
and in particular in \ADFL\ADFT.

\subsec{Effective supergravity}
Let us first recall briefly the $\cx N=2$ effective supergravity 
for the bulk fields of the orientifold \ofi\ from ref.\ADFL. There are 4 vector
fields from reducing the ten-dimensional 2-forms $B_{NS}$ and $B_{RR}$
with one leg along $T^2$. 
Three of them become part of 4d vector multiplets with scalar
components 
\eqn\closedmod{\eqalign{
S&=C_0+i\ e^{-\phi_{10}}\ ,\cr
T&=\int_{K3} C_4+i\ e^{-\phi_{10}}\ Vol(K3)\ ,\cr
U&=\fc{1}{G_{11}}\ (\ G_{12}+i\ \sqrt{\det G}\ )\ ,}}
where $G_{mn}$ is the metric on $T^2/\ZZ_2$, $C_p$ denotes the 
RR $p$-form and $S$ is the complex type IIB dilaton.\foot{$Vol(K3)$
denotes the volume of K3 in the string frame.}
The bulk scalars parametrize locally the special K\"ahler manifold
\eqn\vmb{
\cx M_V^0=\biggl(SU(1,1)/U(1)\biggr)_{ S , U , T }^3.
}
The hyper multiplet moduli space 
\eqn\mh{
{\cal M}_H={O(\Gamma^{4,20})\backslash SO(4,20)/SO(4)\times SO(20)}\ ,}
is associated with 
the moduli of $K3$ up to one subtlety. In the decomposition
\eqn\ktdec{
SO(4,20)\simeq SO(3,19) \times \bx R^{22} \times \bx R_+ ,
}
the coset associated with the first factor describes the Einstein metrics of 
constant volume discussed before and the second factor
the $B$-fields on $H^2(K3)$. However, the last factor describes
the K\"ahler volume of $T^2$ \ADFL.

One may readily identify a $(SU(1,1)/U(1))_{ S , U }^2$ factor
parametrized by the 8d fields $ S $ and $ U $ in $\cx M_V^0$
as the subset \ett\ of the F-theory compactification. However the 
full 8d F-theory spectrum, including the brane 
gauge fields and the moduli scalars $C^a$, 
is given by the complex structure moduli in \modellk. The enlarged
special K\"ahler manifold that includes the three bulk fields \vmb\ as 
well as the brane degrees of freedom is therefore%
\foot{Another 
quick way to get the following moduli space is to use the 8-dimensional
duality \vafaf\ between F-theory on K3 and the heterotic string on $T^2$,
and the moduli space for the perturbative 
heterotic string compactified on $T^2\times$ K3 \hetsug.} 
\eqn\msugra{
\cx M_V=\fc{SU(1,1)}{U(1)} \times \fc{SO(2,18)}{SO(2)\times SO(18)}.}
A similar proposal, including also possible moduli from D3 branes, 
has been made in \ADFT\  based on constraints from electric-magnetic 
duality. However as discussed around eq.(4.10), the F-theory 
derivation adds some important extra physical information related to the
integral lattice of BPS states and the 
D7-brane positions.

\subsec{Supergravity gaugings vs. flux potentials}
We will now give the precise dictionary between the F-theory fluxes
in the superpotential \supiii\ and the gaugings in the effective supergravity.
We will use the notations of \thebible, to which we also refer 
for background material on gauged supergravities.

In the formalism of $\cx N=2$ supergravity, the scalar potential,
for purely electric gauging, is given by the expression
$$
V=4h_{uv}k^u_\Lambda k^v_\Sigma L^\Lambda\bar L^\Sigma+
(U^{\Lambda\Sigma}-3\bar L^\Lambda
L^\Sigma)P_\Lambda^xP_\Sigma^x,
$$%
where%
$$U^{\Lambda\Sigma}=-\frac 1 2
(\Im\cx N_{\Lambda\Sigma})^{-1}-\bar L^\Lambda L^\Sigma,
\qquad L^\Lambda=e^{K_V/2}\, X^\Lambda.
$$
Here $(X^\La,F_\La)$, $\La=0,...,n_V$  
denotes the symplectic section that defines the special geometry of the 
$n_V$ vector multiplets,
$\cx N_{\Lambda\Sigma}$ is the coupling matrix of the gauge fields and
$K_V$ is the K\"ahler potential on $\cx M_V$. Moreover $h_{uv}$ is
the metric for the hyper multiplets. The couplings between hyper multiplets
and vector multiplets are described by the Killing vectors  $k^u_\La$ and the
associated prepotentials $P^x_\La$. The index $x$ labels a 
triplet under the $SU(2)$ symmetry, which is embedded in the 
holonomy of the quaternionic manifold $\cx M_H$.

Under certain conditions, the above scalar potential can be related
to the scalar potential of an $\cx N=1$ supergravity obtained
from $F$-terms and $D$-terms. To do so, one can try to 
split the contributions to $V$ into a candidate $F$-term 
related to, say,  the complex prepotentials
\eqn\defPc{P_\La=P^1_\La+iP^2_\La,} and a $D$-term related to $P_\La^3$,
or another choice related to it by $SU(2)$ rotations. 
The candidate expression for the holomorphic superpotential of $\cx N=1$
supergravity for the flux potential \sp\ is then \PMsp\cupl
\eqn\wsug{
W = \sum_\La P'_\La X^\La-\sum_\La \tilde{P}^{\prime\La} F_\La,
}
where the tilded prepotentials $\tilde{P^\La}$
describe the coupling of the dual magnetic fields and the
prime denotes the holomorphic sections, related to $P_\La$ by the
standard exponential in the K\"ahler potential.
The ansatz \wsug\ is justified, if the scalar
potential $V$ can be written as a standard $\cx N=1$ potential based on 
$W$ and an additional $D$-term that depends on the real prepotential
$P^3_\La$.

\subsubsec{Dependence on $\cx M_H$}
The manifold $\cx M_H$ has translational isometries
corresponding to shifts of the 22 $B$-fields on K3,
made explicit in the decomposition \ktdec. 
For these ``axionic'' gaugings there is a simple 
relation between the prepotentials and the Killing vectors 
\eqn\kprep{
P^x_\La = \om_I^xk^I_\La,
}
where $\om_I^x$ is the $SU(2)$ connection \IIBsugra.
It will turn out,
that the objects $\om_I^x$ are essentially the periods $\Pi_I^x$ defined 
in \defkt, up to an overall factor. One has 
$$
\om^x_I \sim e^\al\, (L^{-1})^x_I,
$$
where $L$ is a coset representative of $SO(3,19)/SO(3)\times SO(19)$
and $\al$ parametrizes the volume factor \ADFL. 

For the purpose of defining the superpotential \wsug, we must 
choose now one direction in $SU(2)$, say $x=3$, and  define the complex
holomorphic prepotential \defPc. In the present case this
is the same as choosing a fixed complex structure on the ``lower'' 
$\kb$ and this is already anticipated by the split \supM\ of the M-theory 
superpotential. As discussed in sect.~3.1, the superpotential $W$ becomes
the 4-dimensional $F-$term while $d\Wh$ will enters 
a $D$-term in the $\cx N=1$ sense.

Accordingly, we introduce the complex quantities
$$
\om_I \equiv \om^0_I+i\om^1_I \sim e^\al L^J M_{JI},\hskip 60pt
L^I=L^I_1+iL^I_2,
$$
where we used $L^T M L=M$. 

Finally, it follows from $L^I M_{IJ} \bb L^J = 1$ that
$
L^I=e^{K_0/2}\, \Pi^I,
$
with $\Pi^I$ the holomorphic period vector and $K_0$ the K\"ahler potential
for the coset based on $SO(3,19)$. The final form for
the complex Killing prepotential \defPc\ is then
\eqn\resP{
P_\La = e^{K_H/2}\ \Pi^I\,  M_{IJ}\,  k^J_\La,
}
where $K_H=K_0+2\al$ is the K\"ahler potential on $\cx M_H$ and
a similar formula holds for the real prepotential $P^3_\La$.

\subsubsec{Dependence on $\cx M_V$}
The defining data of the effective supergravity for the $n_V$ 
is the symplectic section $(X^\La,F_\La)$. As explained
in \ADFT, the geometry of the 8-dimensional vector multiplets 
is unaffected by the compactification on the ``lower'' K3, $\kb$.
This implies that the symplectic section is given by a 
product of the 8-dimensional period vector $\Pit_\La$ in \pvcs\ and the 
period vector $(1, T )$ for the volume of $\kb$:%
\foot{%
To make contact with the standard  supergravity conventions, 
one must transform the period vector $\Pit^\La$ 
into the orthonormal basis (A.7).}
\eqn\defXFc{
X^\La=\pmatrix{1\cr- S  U +\h C^aC^a \cr S \cr U \cr C^a}
,\qquad F_\La =  T\  M_{\La\Si}\ \Pit^\Si.
}
Note that the F-theory derivation of the section $X^\La=\Pit^\La$, as the
period vector of the ``upper'' K3, provides important physical information
that does not follow from the duality argument in supergravity used in \ADFT. 
Specifically the period vector is defined with reference to the 
{\it integral} lattice $H^2(\kf,\ZZ)$, which
represents the lattice of BPS charges arising from the 20 8-dimensional 
$U(1)$ gauge fields. In this context, $X^\La$ represents the 
central charge of  4d BPS domain walls carrying a certain 
gauge charge.

Without reference to the integral lattice, the supergravity section 
$X^\La$ cannot determine the D7-brane positions. In fact any point
on $\cx M_V$ can be parametrized locally by a section \defXFc\ with 
$C^a=0$ and $ S , U $ some complex numbers. E.g., the different physics
of D7-branes at generic positions on one side,
and  D7-branes on top of the orientifold plane on the other side,  
is described by the different sets of integral vectors orthogonal to
the section $X^\La$ at, say, $C^a=0$.
Similarly the integrality properties of the flux, or the gaugings in 
supergravity, refer to a symplectic section in the integral basis,
which is provided by the geometry of the F-theory compactification.

\subsubsec{Mapping fluxes and gaugings}
Combining the above equations, the candidate superpotential 
becomes a degree (1,1) polynomial in the periods
of the two K3's:
\eqn\resWsug{
W=\sum_\La P'_\La X^\La=
\sum_{I\La}\ \Pi^I\,  \Pit^\Si\, k_{I\La},}
where we removed the exponential $e^{K_H/2}$ to pass to the
holomorphic section $P'_\La$ and $k_{I\La}=M_{IJ}k^I_{\ \La}$. 
Note that $W$ is also 
an inhomogeneous polynomial of degree 4 in the 
K3 moduli. 

The expression \resWsug\ is in perfect agreement with \supii\ if
we identify the F-theory fluxes \defgf\  and the Killing vectors 
$k^I_\La$ as 
\eqn\resc{
\vbox{\offinterlineskip\tabskip=0pt\halign{\strut
\vrule\ \ \ \hfil~$#$~\hfil\ \ \ \vrule\cr
\noalign{\hrule}
\cr
\g^I_{\ \La}=k^I_{\ \La}.\cr
\cr
\noalign{\hrule}}}}
As already explained in detail in sect.~2.5., near the orientifold,
the fluxes with index $\La=1,...4$ are identified with the bulk 
3-form flux $G_3$, while the fluxes with $\La=5,...,16$ describe the
2-form flux on the D7-brane. At the risk of being repetitive, 
let us summarize this relation between fluxes and Killing
vectors near the orientifold limit\foot{The following identifications refer
to the basis \pvcs\ of the period vector, defined in terms of the
integral basis $\{\eta^I\}$ for $H^2(\kb,\ZZ)$; moreover $dz=dx+ U  dy$
is the complex coordinate on $T^2/\ZZ_2$, as before.}

\eqn\kds{
\vbox{\offinterlineskip\tabskip=0pt\halign{\strut
\ \hfil~$#$~\ = \  &~$#$~\hskip40pt\hfil &~#~\hfil\cr
k^I_1& U _2^{-1}\int F_3\wedge(\eta^Idx)&RR bulk 3-form\cr
k^I_4& U _2^{-1}\int F_3\wedge(\eta^Idy)&\hskip28pt  '' \cr
k^I_2& U _2^{-1}\int H_3\wedge(\eta^Idy)&NS bulk 3-form\cr
k^I_3&\hskip-8pt - U _2^{-1}\int H_3\wedge(\eta^Idx)&\hskip28pt ''\cr
k^I_{a}&\hskip18pt\int F_2^a\wedge \eta^I&16 D7-brane 2-forms $(a>4)$\cr
}}
}
The supergravity prepotentials $P^x_\La$,
associated with these fluxes are obtained
by contracting these vectors with the K3 periods as in \kprep.

\subsec{Vacua in gauged supergravity}
It is remarkable that the analysis
of the vacuum configurations is much simpler in the flux language
then in the supergravity language. As an illustration of this point, 
we use the identification \resc\ to display 
``new'' vacua of the effective gauged supergravity.

In fact it has been known for a while that it is quite
hard to find $\cx N=1$ supersymmetric vacua by gauging 
a $\cx N=2$ supergravity. There is a no-go theorem \CGP\ that 
says that this is impossible if the section $(X^\La,F_\La)$
can be derived from a prepotential $\cx F$ such that
$F_\La=\p \cx F/\p X^\La$. The non-existence of a prepotential 
is a very strong condition
which requires a special scalar manifold $\cx M_V$ and in addition
a special choice for the symplectic section $(X^\La,F_\La)$ \FGP.
Remarkably enough, the above map from the flux potential to a
gauged supergravity has precisely led to a special 
section \defXFc\ that can not be derived from a prepotential.
This is consistent with the existence of  $\cx N=1$ supersymmetric
vacua, reported in \TT\ and sect.~3. 

As a simple example let us first reconsider the bulk moduli of the orientifold.
The 4-dimensional slice \ett\ of the orientifold has already been study
from the two different perspectives, namely 
using the superpotential \sp\ in \TT\
and using gauged supergravity in \ADFL, however without an explicit
comparison. The flux superpotential obtained from \sp\ in \TT\ is
\eqn\suptt{
W\sim\int \om \wedge (-\al_y+ S \, \be_y+ U \, \al_x - S \,  U \, \be_x)
\equiv \int \om \wedge \mu,
}
where $\al_a,\ \be_a\in H^2(K3,\ZZ)$ are defined by the 3-form fluxes
$$
F_3 = \al_x dx +\al_y dy,\qquad H_3 = \be_x dx +\be_y dy.
$$
On the other hand, the supergravity potential is obtained from
\wsug\ 
with the period vector \pvcs\ transformed to the orthogonal basis
with metric diag$(1,1,-1,-1)$
$$
X^\La=\fc{1}{\sqrt{2}}\pmatrix{1- S   U \cr- S - U \cr-(1+ S   U )\cr
 U - S }=\fc{1}{\sqrt{2}}\pmatrix{1&1&0&0\cr 0&0&-1&-1\cr 1&-1&0&0\cr
0&0&-1&1}\ \Pit|_{C^a=0}.
$$
The explicit mapping \resc\ from fluxes to Killing vectors $k^I_\La$
of the gauged  supergravity is\foot{In this formula we use the 
supergravity notation of \ADFL\ with $\La=0,...,3$.}:
\eqn\ids{\eqalign{
\al_x=\fc{1}{\sqrt{2}}(-k^I_1+k^I_3)\, \eta_I,\qquad
&\al_y=\fc{1}{\sqrt{2}}(-k^I_0+k^I_2)\,\eta_I,\cr
\be_x=\fc{1}{\sqrt{2}}(k^I_0+k^I_2)\,\eta_I,\ \ \ \qquad
&\be_y=\fc{1}{\sqrt{2}}(-k^I_1-k^I_3)\,\eta_I,\cr
}}
The condition $G_1=0$ on the $\cx N=2$ supersymmetric branch corresponds to
the vanishing of the 2-form $\mu$ in \suptt; this gives
two real condition on the four component 2-forms $\al_a,\ \be_a$,
or two conditions on the four Killing vectors $k^I_\La$. E.g., 
a choice of fluxes that fixes $ S $ and $ U $ at purely imaginary
values is
$$
k^I_0=\fc{1- S _2 U _2}{1+ S _2 U _2}\, k^I_2\, ,
\qquad k^I_1=\fc{ U _2- S _2}{ U _2+ S _2}\, k^I_3\, ,
$$
where $ S = S _1+i\,  S _2$ and similarly for $ U $. 
More generally, scanning through complex vev's for $ S $ and $ U $ one obtains 
a family parametrized by four real parameters that contains
the special $\cx N=2$ supersymmetric solution found in \ADFL.

\subsubsec{Gauged supergravity vacua with 2-form fluxes}
In \ADFT, duality arguments have been used to propose 
an effective supergravity description 
for the type IIB string on \ofi, including D7- and D3-brane moduli. 
In the absence of D3-branes and setting $n_V=18$, this manifold
reduces to the 4d F-theory moduli space \msugra. 

As discussed already, F-theory provides extra physical information, such as 
the integral structure defined by the lattice of BPS states.
Moreover it gives a string theory interpretation of the extra 16 Killing 
prepotentials in \kds, as 2-form fluxes on the 16 D7-branes. 
The vacuum structure for special choice of Killing vectors has also been
studied in \ADFT, with the result that the D7-branes are always
confined to the origin $C^a=0$. This is somewhat at odds with the 
claim of sect.~3, that the generic flux fixes the D7-branes at generic 
position $C^a\neq 0$.

However, using the flux-gauging dictionary in \kds,  it is straightforward to check  
that the choice of Killing vectors in \ADFT\ corresponds to the
special  situation, where 
the plane associated with the 2-form flux $\FF$ is orthogonal to the plane
$\HH$ defined by the bulk flux. Thus the vacua of \ADFT\ fit into the classification
of the orientifold preserving fluxes in sect.~3.3, in agreement with
the fact that $C^a=0$ at the minimum.

\newsec{Superpotential and  open string moduli in orbifold models}
In this and the next section we study in depth the \tb orientifold 
$K3\times T^2/\Om$ in the case where the K3 is the orbifold $T^4/\ZZ_2$.
In particular we describe the full superpotential including open string moduli 
and 2-form fluxes and derive the 
effective gauged  supergravity describing the orbifold.

\subsec{The \tb $K3\times T^2/\Om$ orientifold}
{\it \hskip-0.75cm Geometrical moduli}

First, consider the case \tb on $K3\times T^2$ without the  orientifold projection.
The corresponding moduli space takes the form:
\eqn\nfour
{{\cal M}={SO(22,6)\over SO(22)\times SO(6)}\otimes \Biggl({SU(1,1)\over
U(1)}\Biggr)_U\ .}
The first factor parametrizes  the $\sigma$-model manifold of 132 scalar fields,
which are the scalar components of 22 N=4 vector supermultiplets.
Among them there are the $57$ K\"ahler and complex structure 
moduli at constant volume of $K3$, the \tb dilaton, the two 
volume moduli of $K3$ and 
$T^2$ plus several other Ramond-Ramond scalar fields. Note that
the modulus $U$, which parametrizes
the second factor in \eqq \nfour\
and which corresponds to the complex structure of $T^2$, is a member
of the N=4 supergravity multiplet.

Now, we introduce the orientifold projection $\Om(-1)^{F_L}I_2^3$, with
$I_2^3$ acting on the complex torus coordinate $z^3$ as: $I_2^3:\ z^3\ra-z^3$.
This gives an N=2 closed string spectrum.
The geometrical N=4 moduli space in \eqq \nfour\ gets truncated and 
factorizes locally into a space of N=2 vector-- times N=2 hypermultiplets
\ADFL:
\eqn\mvmh{
\Mc={\cal M}^0_V\otimes {\cal M}_H\ ,}
with $\cx M^0_V$ the special K\"ahler manifold for the moduli \closedmod\ given in \vmb.
The metric of these fields is described by the K\"ahler potential
\eqn\kaehler{
K_V=-\ln (S-\ov S)(T-\ov T)(U-\ov U)\ ,}
which is derived from the following N=2 prepotential:
\eqn\prep{
\Fc(S,T,U)=STU\ .}

The geometric moduli space of $K3$ is described by the coset space 
$$\fc{SO(19,3)}{SO(19)\times SO(3)\times SO(19,3,\IZ)}\times \IR^+$$ 
with real dimensions $58$, while the stringy moduli space of $K3$ 
is given by the coset in \mh, except for the exchange of volume factors from $T^2$ and
$K3$ \ADFL.
The additional $22$ scalars 
 represent the $22$ axions $(b^m,b^a)\ ,\ m=1,2,3,\ a=1,\ldots,19$.
They
originate from reducing the $RR$ $4$--form on $K3\times T^2$:
Due to $h_{(2,0)}(K3)=h_{(0,2)}(K3)=1$ and $h_{(1,1)}(K3)=20$, we obtain the two fields
$C_{mn 3\ov 3}$ and $C_{\ov m\ov n 3\ov 3}$ and the 20 fields $C_{m\ov n 3 \ov 3}$
with the index $3$ referring to the $(1,1)$--form of the torus $T^{2,3}$.
Note, that in $D=4$ these scalars are dual to the anti--symmetric tensors 
$C_{mn \mu\nu}$, $C_{\ov m\ov n \mu\nu}$ and the 20 fields $C_{m\ov n \mu\nu}$, 
respectively. 
The latter are eliminated as a result of imposing self--duality
on the self--dual $4$--form $C_4$. At any rate, we thus obtain $22$ scalars from the $4$--form
$C_4$ reduced w.r.t. the cohomology $H^2(K3)$.
The $80$ $K3$--moduli appear as scalars of $20$ N=2 hypermultiplets.
Furthermore, due to $h_{(0,0)}(K3)=h_{(2,2)}(K3)=1$
we obtain the field $C_{\mu\nu\rho\sigma}$, which gives rise to a tadpole in $D=4$,
and  $a^3=\int_{K3} C_{m \ov n p \ov q}$ 
representing the axion of the K\"ahler modulus of the torus $T^{2,3}$. 
The latter is dual to the 
anti--symmetric tensor $\int_{T^2} C_{\mu\nu 3\ov 3}$, which is eliminated after 
imposing self--duality on $C_4$.

\br
{\hskip-0.75cm \it Adding D3- and D7-branes}

To cancel the tadpoles we add a system of $n_3$ space--time filling $D3$--branes and $n_7$ 
$D7$--branes wrapped on $K3$.  On the internal $D7$--brane world volume
non--trivial two--form gauge fluxes $\cx F^a$ may be turned on. 
The open string $D7$--brane moduli consist of the transverse positions 
$C^{7,a}:=C^{7,a}_3$ of the $n_7$ $D7$--branes along the torus $T^2$ $(a=1,\ldots,n_7)$ and 
$3 n_3$ $D3$--brane moduli $C^{3,b}_i$ accounting for the transverse
$D3$--brane positions along $K3\times T^2$ $(b=1,\ldots,n_3)$.
The scalars $C^{3,b}_i\ ,\ i=1,2$ give rise to additional hypermultiplet scalars.
On the other hand, 
in addition to the three fields $S,T,U$ we have the $n_7+n_3$ scalars 
$C^{7,a},C^{3,b}:=C^{3,b}_3$ of
N=2 vectormultiplets in $D=4$.
In the following, we shall disregard the open string moduli $C_i^{3,b}\ ,\ i=1,2$
giving rise to additional hypermultiplet scalars. 

The metric of these vectormultiplet fields is 
described by the K\"ahler potential 
\threeref\ABFPT\ADFT\JL:
\eqn\kaehler{\eqalign{
K_V&=-\ln\lf[(S-\ov S)(T-\ov T)(U-\ov U)-\h\ (T-\ov T)\ \sum_{a=1}^{n_7} 
(C^{7,a}-\ov C^{7,a})^2\ri.\cr
&\lf.-\h\ (S-\ov S)\ \sum_{b=1}^{n_3} (C^{3,b}-\ov C^{3,b})^2\ri]\ ,}}
which is derived from the following N=2 prepotential \ADFT:
\eqn\prepp{
\Fc(S,T,U,C^{7,a},C^{3,b})=STU-\h\ T\ \sum_{a=1}^{n_7} (C^{7,a})^2-
\h\ S\ \sum_{b=1}^{n_3} (C^{3,b})^2\ .}

On the internal $D7$--brane world volume
the non--trivial (magnetic) two--form gauge flux 
\eqn\twoflux{
\Fc^a_{NP}=\Fc^a_{12}+\Fc^a_{34}:=2\pi\ap\ \lf(\ F^a_{12}\ dx^1\wedge dy^1+F^a_{34}\ 
dx^2\wedge dy^2\ \ri)} 
may be turned on. The latter obey the quantization rule $F^a_{ij}=2\pi\fc{n_{ij}^a}{m_{ij}^a}$, 
\ie: 
\eqn\quant{
f^a_{ij}=\fc{1}{(2\pi)^2}\int_{C_{ij}} \Fc^a_{ij}=\ap\ \fc{n^a_{ij}}{m^a_{ij}}\ \ \ ,\ \ \ 
(i,j)=(1,2)\ ,\ (3,4)\ .}
The dependence of the moduli and matter field metrics on these $2$--form fluxes 
has been derived in \doubref\LRSi\LMRS:
\eqn\results{\eqalign{
G_{T\ov T}&=-\fc{1}{(T-\ov T)^2}\ \ \ ,\ \ \ G_{S\ov S}=-\fc{1}{(S-\ov S)^2}\ \ \ ,\ \ \ 
G_{U\ov U}=-\fc{1}{(U-\ov U)^2}\ ,\cr
G_{C^{7,a}\ov C^{7,a}}&=-\fc{1}{(S-\ov S)(U-\ov U)}+
\fc{\ap^{-2}\ f^a_{12} f^a_{34}}{(T-\ov T)(U-\ov U)}\ ,\cr
G_{C^{3,b}\ov C^{3,b}}&=\fc{-1}{(T-\ov T)(U-\ov U)}\ .}}
In fact, up to second order in the matter fields $C^{7,a},\ C^{3,b}$,
we may summarize these results in the K\"ahler potential 
\eqn\kaehler{\eqalign{
K_V&=-\ln\lf[(S-\ov S)(T-\ov T)(U-\ov U)\ri.\cr
&-\h\ \sum_{a=1}^{n_7}\ [\ (T-\ov T)-(S-\ov S)\ f_{12}^af_{34}^a\ \ap^{-2}\ ]\ 
(C^{7,a}-\ov C^{7,a})^2\ ,\cr
&\lf.-\h\ \sum_{b=1}^{n_3} (S-\ov S)\ (C^{3,b}-\ov C^{3,b})^2 \ri]\ ,}}
which is derived from the following N=2 prepotential:
\eqn\prep{
\Fc(S,T,U,C^{7,a},C^{3,b})=STU-\h\ \sum_{a=1}^{n_7} (T-\ap^{-2}\ S\ f_{12}^af_{34}^a)\ (C^{7,a})^2-
\h\ S\ \sum_{b=1}^{n_3} (C^{3,b})^2\ .}
It may be interesting to note, that a similar instanton--number dependent prepotential
arises in heterotic $K3\times T^2$ compactifications \stieberg.

In order to preserve 
supersymmetry on the $D7$--brane--world volume, the $2$--form fluxes 
have to obey \AAS
\eqn\susy{
\fc{f^a_{12}}{\im T^2}=-\fc{f^a_{34}}{\im T^1}\ .}
Note that these equations are nothing but the $D$-term equation \sdtof.
The two fluxes $f^a_{12},f^a_{34}$ are quantized according to \quant. This fixes 
the ratio $\im T^1/\im T^2$ to
\eqn\fixed{
\fc{\im T^1}{\im T^2}=-\fc{n^a_{34}\ m^a_{12}}{n^a_{12}\ m^a_{34}}\ .}
Later we shall see, that these two--form fluxes correspond to a specific gauging
in the underlying low--energy supergravity description, which fixes the ratio $\im T^1/\im T_2$.

\subsec{The moduli space in the orbifold limit $T^4/\ZZ_2$ of $K3$}

In the following, let us work in an  orbifold limit of $K3$. In the case of a 
$\IZ_2$--orbifold limit the hypermultiplet moduli space $\Mc_{H}$ in \mh\
boils down to the coset
\eqn\HYPERB{
\Mc_H=\fc{SO(4,4)}{SO(4)\times SO(4)\times SO(4,4,\IZ)}\ ,}
whose real dimension is $16$. 
The coset may be described  by the symmetric $G$ and anti--symmetric 
background $B$, with $G^t=G$ and $B^t=-B$.
More precisely, in Narain lattice compactifications one 
introduces the moduli scalar matrix\foot{Note, that alternatively, by the 
unimodular transformation $\fc{1}{\sqrt 2}\pmatrix{{\bf 1}_d& {\bf 1}_d\cr
-{\bf 1}_d& {\bf 1}_d}$ one may transform the background $\Mc$ into the form:
$$\Mc=\pmatrix{G^{-1}&G^{-1}B\cr -B G^{-1}& G-BG^{-1}G}\ .$$}
\eqn\MSM{
\Mc=\h\pmatrix{G+G^{-1}+G^{-1}B-BG^{-1}-BG^{-1}B && G-G^{-1}+G^{-1}B+BG^{-1}-BG^{-1}B \cr
             && \cr
G-G^{-1}-G^{-1}B-BG^{-1}-BG^{-1}B && G+G^{-1}-G^{-1}B+BG^{-1}-BG^{-1}B }}
describing a general $\fc{SO(d,d)}{SO(d)\times SO(d)}$--background.
On may verify $\Mc^t=\Mc$, $\det(\Mc)=1$ and $\Mc\eta \Mc^t=\eta$, with 
$\eta=diag(+1^d,-1^d)$.
The moduli metric is given by $ds^2=-\fc{1}{4}\ \Tr(d\Mc\ d\Mc^{-1})$.

For our case, \ie $d=4$, the 
background $G$ and $B$ gives rise to $16$ real parameters, which comprise the 
two complex structure moduli $U^1,U^2$ and six K\"ahler moduli 
$\Tc^i\ ,\ i=1,\ldots,6$ of $T^4/\IZ_2$. 
Note that the split into K\"ahler and complex structure
moduli is only a convention on $K3$. The K\"ahler moduli $\Tc^i$ are complexified 
with the $RR$--axions $a^i$ encoded in the anti--symmetric background:
\eqn\bfield{
B=\pmatrix{0&a^1 & a^3 &a^4\cr
           -a^1&0&a^5&a^6\cr
           -a^3&-a^5&0&a^2\cr
           -a^4&-a^6&-a^2&0}\ .}
For the case under consideration \HYPERB\ the decomposition \ktdec\ translates into:
\eqn\decomp{
\fc{SO(4,4)}{SO(4)\times SO(4)}=\fc{SO(3,3)}{SO(3)\times SO(3)}\times \IR_+ \times\ \IR^6\ .}
The coset $\fc{SO(3,3)}{SO(3)\times SO(3)}$ describes nine of the ten parameter 
of $G$, while the factor $\IR_+$ describes the volume
of $T^4$. In the next section
 we shall discuss the gauging of the remaining six PQ--symmetries 
of $(b^m,b^a)\ ,\ m,a=1,2,3$, with the following identifications:
\eqn\identification{
\lf(b^m\atop b^a\ri)=\fc{1}{\sqrt 2}\ \pmatrix{-a^4-a^5 \cr 
                                                 a^3-a^6 \cr 
                                                 a^1+a^2 \cr 
                                                 a^4-a^5 \cr
                                                 a^3+a^6 \cr
                                                -a^1+a^2}
\qquad \ \ \  m,a=1,2,3 .}

Generically, the metric of the space \HYPERB\ is given by:
\eqn\Metricc{
ds_{SO(4,4)}^2=-\h\Tr(dG^{-1}\ dG)-\h\Tr(G^{-1}\ dB\ G^{-1}\ dB)\ .}
If we specialize the torus $T^4$
to be a direct product of two two--tori $T^4\simeq T^{2,1}\times T^{2,2}$, with
the {\it field--theoretical} moduli\foot{We have the following relations between physical 
and geometric quantities \doubref\LMRS\LRSi:
\eqn\relm{
\im T^i=\fc{1}{2\pi\ap^2}e^{-\phi_{10}} \Tc_2^j \Tc_2^k=
\fc{1}{2\pi\ap^{1/2}}\ e^{-\phi_4}\ \sqrt\fc{\Tc_2^j\Tc_2^k}{\Tc_2^i}\ .}
Furthermore, the $K3$-- or $T^4$--volume is given by
$Vol(K3)=(2\pi\ap^2)^{-1}\Tc^1_2\Tc^2_2$, and the torus volume 
$Vol(T^2)=\Tc_2^3$. In addition, we have: $\phi_{10}=\phi_4+
\h\ln(\Tc_2^1\Tc_2^2\Tc_2^3/\ap^3)$.}
\eqn\moduli{
T^j=a^j+i\ \sqrt{\det G^j}\ \ \ ,\ \ \ U^j=\fc{1}{G_{11}^j}\ (\ G_{12}^j+i\ \sqrt{\det G^j}\ )
\ \ \ ,\ \ \ j=1,2\ ,}
respectively, the background $G$ and $B$ in \Metricc\ may be parametrized by
\eqn\introduce{
B:=\pmatrix{B_1&B_3\cr  -B_3^t&B_2}\ \ \ ,\ \ \ G:=\pmatrix{G_1&0\cr 0&G_2}\ ,}
with the $2\times 2$ matrices $B_i\ ,\ i=1,2,3$ and $G_1,G_2$: 
\eqn\with{
G_i=\fc{T_2^i}{U_2^i}\ \pmatrix{1&U_1^i\cr U_1^i& |U^i|^2}\ \ \ ,\ \ \ i=1,2\ .}
In \eqq \introduce\ we have kept all six RR--scalars $a^j$ 
in order to discuss the gauging of all six isometries of \identification. 
With the moduli $T^j,U^j\ ,\ j=1,2$, the metric may be also written:
\eqn\alsowr{
ds_{SO(4,4)}^2=4\sum_{M,N\in\{T^i,U^j\}}
\fc{\p^2K_H}{\p M\p \ov N}+\Tr(G_1^{-1} dB_3\ G_2^{-1} dB_3)\ ,}
with the K\"ahler potential:
\eqn\hypermetric{
K_H=-\sum_{j=1}^2\ln(U^j-\ov U^j)-\sum_{j=1}^2\ln(T^j-\ov T^j)\ .}

In the following we shall introduce new fields such, that in \eqq \Metricc\ and \alsowr\ 
the decomposition \decomp\ becomes visible. This is achieved through two steps.
We first  split off the volume factor $\IR_+$ in \decomp\ by introducing the 
new fields
\eqn\fields{\eqalign{
\rho&=\fc{\im(T^1)}{\im(T^2)}=\fc{\im\Tc^2}{\im\Tc^1}\ ,\cr
e^\Phi&=\sqrt{\im(T^1)\im(T^2)}=
\fc{e^{-\phi_{10}}}{2\pi\ap^2}\ \sqrt{\Tc_2^1\Tc_2^2}\Tc_2^3=\fc{e^{-\phi_4}}{2\pi\ap^{1/2}}
\ \sqrt{\Tc_2^3}\ ,}}
with $e^{2\Phi}$ essentially 
describing the volume $Vol(T^2)$ of the torus $T^2$,
\ie the factor $\IR_+$ in \decomp. 
With these new coordinates the metric \Metricc\ of the coset \HYPERB\ may be  written:
\eqn\metric{\eqalign{
ds^2_{SO(4,4)}&=-4\fc{dU^1\ d\ov U^1}{(U^1-\ov U^1)^2}-4\fc{dU^2\ d\ov U^2}{(U^2-\ov U^2)^2}
+\h\ \fc{d\rho^2}{\rho^2}+2\ d\Phi^2\cr
&+e^{-2\Phi}\ \lf[\ \fc{(da^1)^2}{\rho}+
\rho\ (da^2)^2\ \ri]+\Tr(G_1^{-1} dB_3\ G_2^{-1} dB_3)\ .}}
As a second step we have to introduce the parametrization of the coset 
$\fc{SO(3,3)}{SO(3)\times SO(3)}$, which accounts for the 
remaining metric moduli $U^1,U^2,\rho$. This has to be achieved  
such, that the metric of the $RR$--scalars $b^m,b^a$ takes
the  appropriate form, dictated by the decomposition \decomp.

In the supergravity literature, see {\it Ref.} \thebible\ 
and in particular the {\it Refs.} \doubref\ADFL\ADFT, 
the local parametrization of the moduli space 
$\fc{SO(4,4)}{SO(4)\times SO(4)}$ according to the decomposition \decomp\ is 
given by the coset representative $e^{C^IZ_I}\ e^{\phi S}\ L$,
with $Z_I$ the generators of the Abelian shift symmetries on $b^I$, $S$ a generator of
$SO(1,1)$ and $L$ being a coset representative of $SO(3,3)$. 

Generically, the coset representative $L$ of the manifold $\fc{SO(d,d)}{SO(d)\times SO(d)}$
is parametrized by the $d^2$ coordinates 
$e=\{e_m^{\ \ a}\}$, $e^t=\{e_a^{\ \ m}\},\ m,a=1,\ldots,d$ 
\eqn\Lcoset{
L_M^{\ \ A}=\pmatrix{\sqrt{1+ee^t}&e \cr   e^t& \sqrt{1+e^te}}\ ,}
with $L\eta L^t=\eta$, implying $L_M^{\ \ A}\eta_{AB}L_N^{\ \ B}=\eta_{MN}$.
Any $SO(d,d)$ background $\Mc$, described by $G$ and $B$ as given in \eqq \MSM, 
may be expressed by 
\eqn\link{
\Mc=L^tL\ ,}
with the matrix representation:
\eqn\coset{
L=\h\pmatrix{q+q^\ast+q^\ast B & &q-q^\ast+q^\ast B \cr
             q-q^\ast-q^\ast B  & & q+q^\ast-q^\ast B  }\ .}
Here, $q$ is the vielbein $q_{ma}$, with $G_{mn}=\delta^{ab}\ q_{ma}q_{nb}$, \ie $G=q q^t$
and $q^\ast=(q^t)^{-1}$.
To relate the {\it matrix representation} \coset\ to the {\it coordinates} \Lcoset, we find
for the parametrization of $SO(3,3)$: 
\eqn\back{\eqalign{
e_m^{\ \ a}\equiv e=\h\ (q-q^\ast+q^\ast B )\ .}}
On may verify, that $\det(L)=1$ and $L\eta L^t=\eta$. The last matrix identity implies
$\fc{1}{4}(q+q^\ast+q^\ast B)(q+q^\ast+q^\ast B)^t-ee^t=1$, which translates into:
$\sqrt{1+ee^t}_m^{\ \ n}\equiv\sqrt{1+ee^t}=\h\ (q+q^\ast+q^\ast B )$.

After this general discussion about the  parametrization of coset space in supergravity, let us
go back to our metric \metric\ to be decomposed according to \decomp.
The parametrization  $L$ of the coset $\fc{SO(3,3)}{SO(3)\times SO(3)}$, accounting for the 
remaining metric moduli $U^1,U^2,\rho$ of $T^4$ may be found as follows.
Since we restrict the torus $T^4$ to a direct product of two
two--tori, the coset $SO(3,3)_{\rho,U^1,U^2}$ decomposes into the product $SO(2,2)_{U^1,U^2}
\times SO(1,1)_\rho$. The $SO(2,2)$--structure
of $e$ is obtained from \back, with the choice: 
$$G=\fc{U_2^2}{U_2^1}\ \pmatrix{1& U_1^1\cr
U_1^1&|U^1|^2}\ \ \ ,\ \ \ B=\pmatrix{0&U_1^2\cr -U_1^2&0}\ ,$$
\ie: 
\eqn\cos{
q_{SO(2,2)}=\sqrt\fc{U_2^2}{U_2^1}\ \pmatrix{1&U_1^1\cr 0&U_2^1}\ .}
Furthermore, with 
\eqn\cosi{
q_{SO(1,1)}=\sqrt\rho\ ,} 
we find:
\eqn\find{
e=\pmatrix{\fc{-U^1+\ov U^1+U^2-\ov U^2}{2\sqrt{U^1-\ov U^1}\sqrt{U^2-\ov U^2}}&
\fc{U^1\ U^2-\ov U^1\ \ov U^2}{2\sqrt{U^1-\ov U^1}\sqrt{U^2-\ov U^2}}&0\cr
&&\cr
\fc{i\ (U^1+\ov U^1-U^2-\ov U^2)}{2\sqrt{U^1-\ov U^1}\sqrt{U^2-\ov U^2}}&
-\fc{i\ (2+U^1\ U^2+\ov U^1\ \ov U^2)}{2\sqrt{U^1-\ov U^1}\sqrt{U^2-\ov U^2}}&0\cr
0&0&\h\ (\sqrt\rho-\fc{1}{\sqrt\rho})}\ .}
Since $SO(2,2)_{U^1,U^2}\simeq SU(1,1)_{U^1}\times SU(1,1)_{U^2}$, we could have also transformed 
$e$ into a complex basis, where the moduli $U^1$ and $U^2$ are decoupled. 
The latter basis is in fact the period vector $\Pi^I$ 
at the orientifold point and the $3\times 3$ matrix $e$ above is 
essentially, up to an overall factor, 
the triplet of period vectors $\Pi^x_I$ \defkt,
restricted to $SO(2,2)\times SO(1,1) \subset SO(3,19)$.

To this end, we find the $SO(4,4)$ metric \metric\ in terms of $L(e),\Phi$ and the Ramond fields
$b^m,b^a$:
\eqn\Metric{
ds^2_{SO(4,4)}=-\fc{1}{4}\ \Tr\lf[\ d(L^tL)\ d(L^{-1}L^\ast)\ \ri]+2\ d\Phi^2
+e^{-2\Phi}\ \lf(db^m\atop db^a\ri)^t\ L^t\ L\ \lf(db^m\atop db^a\ri)\ .}
Note, that the modulus
$\Phi$, introduced in \eqq \fields, plays the r\^ole of the extra factor 
${\bf R}_+\equiv SO(1,1)$ in the decomposition \decomp.

\subsec{Superpotential from gauged supergravity}

Turning on fluxes in string theory may be understood from the supergravity point of view as
gauging some of the isometries of the moduli space of the scalars under some of the 
gauge fields $A^\Lambda_\mu$ of the vectormultiplets $\Lambda=0,1,2,\ldots,n_3+n_7+3$. 
More precisely,
the relevant isometries in the case under consideration are associated to 
Peccei--Quinn--symmetries of the scalars $(b^m,b^a)$, 
which originate from the $C_4$ $RR$--form:
\eqn\gaugings{\eqalign{
D_\mu b^m&=\p_\mu b^m+f^m_\Lambda\ A^\Lambda_\mu\ ,\cr
D_\mu b^a&=\p_\mu b^a+h^a_\Lambda\ A^\Lambda_\mu\ .}}
In \tb string theory, a subset of these couplings $f^m_\Lambda,h^a_\Lambda$ 
correspond to non--vanishing $NS$ and $R$ three--form fluxes and non--trivial two--form
fluxes on the $D7$--brane world--volume. In the above we have used
the following notations of {\it Ref.} \ADFT\ for the Killing vectors $k^I_{\ \La}$,
$$
k^I_{\ \La}=\cases{
f^m_\La&$I=m=1,2,3$\cr
h^a_\La&$I-2=a=1,2,3$\cr},
$$
which distinguish between directions of different signature.
The conformal Killing potentials 
$P_\Lambda^x\ ,\ x=1,2,3$, are expressed  in {\it Ref.} \ADFT\ as
\eqn\killing{
P_\Lambda=\sqrt 2\ e^{-\Phi}\ [\ (L^{-1})^x_{\ m}\ f^m_\Lambda+
(L^{-1})^x_{\ a}\ h^a_\Lambda\ ]=\sqrt 2\ e^{-\Phi}\ L^{-1}\ \lf(f^m_\Lambda\atop 
h^a_\Lambda\ri)\ .}
With our parametrization  \find\ we obtain:
\eqn\findkilling{\eqalign{
P_\Lambda^1&=\fc{e^{-\Phi}}{\sqrt{2\ (U^1-\ov U^1)(U^2-\ov U^2)}}\ \lf[\ 
f^1_\Lambda\ (U^1-\ov U^1+U^2-\ov U^2)+h^1_\Lambda\ (U^1-\ov U^1-U^2+\ov U^2)\ri.\cr
&\hskip5cm\lf.+(f^2_\Lambda-h^2_\Lambda)\ (U^1U^2-\ov U^1\ov U^2)\ \ri]\ ,\cr
P_\Lambda^2&=-\fc{i\ e^{-\Phi}}{\sqrt{2\ (U^1-\ov U^1)(U^2-\ov U^2)}}\ \lf[\ 
f^1_\Lambda\ (U^1+\ov U^1+U^2+\ov U^2)+h^1_\Lambda\ (U^1+\ov U^1-U^2-\ov U^2)\ri.\cr
&\hskip5cm\lf.+f^2_\Lambda\ (U^1U^2+\ov U^1\ov U^2-2)-h^2_\Lambda\ (U^1U^2+\ov U^1\ov U^2+2)\ \ri]
\ ,\cr
P_\Lambda^3&=\fc{1}{\sqrt 2}\ e^{-\Phi} \lf[\ (f^3_\Lambda+h^3_\Lambda)\ \fc{1}{\sqrt{\rho}}+
(f^3_\Lambda-h^3_\Lambda)\ \sqrt{\rho}\ \ri]\ .}}
The (electrical part of) N=1 superpotential $W$, \eqq \wsug, is:
\eqn\superi{
W=2^{-3/2}\ e^{-K_H/2}\ X^\Lambda\ (P^1+iP^2)_\Lambda\ .}
Here, the parameter $f^m_\Lambda,\ h^a_\Lambda$ refer to the electrical gaugings, 
introduced in \eqq \gaugings.
Furthermore, the symplectic section $(X^\Lambda,F_\Lambda)$ in the orthonormal basis, 
with $X^\Lambda$ given by
\eqn\symsection{\eqalign{
X^0&=\fc{1}{\sqrt 2}\ \lf[1-SU+\h\sum_{a=1}^{n_7} (C^{7,a})^2\ri]\ ,\cr
X^1&=-\fc{1}{\sqrt 2}\ (S+U)\ ,\cr
X^2&=-\fc{1}{\sqrt 2}\ \lf[1+SU-\h\sum_{a=1}^{n_7} (C^{7,a})^2\ri]\ ,\cr
X^3&=-\fc{1}{\sqrt 2}\ (U-S)\ ,\cr
X^a&=C^{7,a}\ ,\cr
X^b&=C^{3,b}\ ,}}
is obtained from \prep\ and the special coordinates  
after a symplectic transformation \ADFT. This 
transformation leads to sections with the property $F_\Lambda\sim T$, which is appropriate for our
$D7$--brane setup.
Note, that that symplectic transformation, given in \ADFT, may be also used for the case
of non--vanishing $f^a_{12},f^a_{34}$ two--form flux components.  As a result, the flux
dependence will only enter the sections $F_\Lambda$.

With the relation $e^{-K_H/2}=2i\ e^{\Phi} \sqrt{(U^1-\ov U^1)\ (U^2-\ov U^2)}$, 
we find for \superi:
\eqn\super{\eqalign{
W(S,T,U^1,U^2,U^3,C^7,C^3)&=\sum_{\Lambda=0}^{4+n_7+n_3}  X^\Lambda\ \lf\{\ f^1_\Lambda\ (U^1+U^2)
-f^2_\Lambda\ (1-U^1\ U^2)\ri.\cr
&\lf.+h^1_\Lambda\ (U^1-U^2)-h^2_\Lambda\ (1+U^1 U^2)\ \ri\}\ ,}}
with the sections $X^\Lambda$ given in \symsection\ and the identification $U\equiv U^3$.
This agrees with the F-theory superpotential \spof\ for $C^{3,b}=0$.
Note, that, in addition, there appear the couplings to the sections $F_\Lambda$ as a matter
of symplectic invariance. The latter are proportional to the $K3$--volume modulus $T$, 
\ie $F_\Lambda\sim T$.

\subsec{Discussion of Superpotential: two--form and three--form fluxes}

In this subsection we shall further explore the relation \resc\ 
between the different directions of gaugings 
$f_\Lambda^m,\ h^a_\Lambda$ and the possible fluxes in the \tb string with $D7$-branes.
The latter is 
compactified on the orbifold $X_6=T^4/\IZ_2\times T^2$ supplemented with the orientifold
projection $\Om(-1)^{F_L}I_2^3$.
For the latter model the $3$--form flux $G_3$ is expressed as linear combination 
of the integral cohomology elements\foot{See appendix \appC\  for more details.} 
$H^3(T^4/\IZ_2\times T^2)$:
\eqn\G{\eqalign{
{1\over{(2\pi)^2\alpha'}}\ G_3&=\sum_{i=0}^3 \lf[\ (a^i-Sc^i)\alpha_i+(b_i-S d_i)\beta^i\ \ri]\cr
&=\sum_{i=0}^3 (\ A^i\omega_{A_i}+B^i\omega_{B_i}\ )\ .}}
The complex 3--form flux is defined as follows: $G_3=F_3-SH_3$, where $F_3=dC_2$ comes from the 
Ramond sector and $H_3=dB_2$ comes from the Neveu--Schwarz sector.
While in the complex basis the Hodge-decomposition $H^3=H^{(3,0)}\oplus H^{(2,1)}\oplus 
H^{(1,2)}\oplus H^{(0,3)}$ of $G_3$ is manifest, the $SL(2,\IZ)_S$--invariance of $G_3$ is 
manifest in the real basis.

Without $D3$-- and $D7$--branes, \ie in the absence of open string moduli and $2$--form
world--volume fluxes on the $D7$--branes, the superpotential originates  from  the
closed string sector only and is given by \sp:
\eqn\TVW{
W={1\over{(2\pi)^2\alpha'}}\ \int_{X_6} G_3\wedge\Omega\ ,}
with $\Omega$ the holomorphic three--form of $X_6$, \ie
$\Omega=dz^1\wedge dz^2\wedge dz^3$.
This is to be compared with \tb compactified on a Calabi--Yau manifold leading to an N=2
closed string sector. It only depends on the dilaton $S$ and complex structure moduli $U^j$.
With the definitions of appendix C, we obtain for \TVW\ the following expression \LRSi:
\eqn\WhT{\eqalign{
W_1(S,U^1,U^2,U^3)&=-(a^1-S\ c^1)\ U^2\ U^3-(a^2-S\ c^2)\ U^1\ U^3-(a^3-S\ c^3)\ U^1\ U^2\cr
&-(b_0-S\ d_0)+(a^0-Sc^0)\ U^1\ U^2\ U^3-\sum_{i=1}^3(b_i-S\ d_i)\ U^i\ .}} 
On the other hand, in the expression \super\ from gauged supergravity, 
that piece \WhT\ is (for $C^{7,a}=0$) captured by the sum $\Lambda=0,\ldots,3$, provided
the following identifications are made:
\eqn\match{\eqalign{
\sqrt{2}\ a^0&=-f^2_1-f^2_3+h^2_1+h^2_3\ \ \ ,\ \ \ \sqrt{2}\ a^1=f^1_1+f^1_3-h^1_1-h^1_3\ ,\cr
\sqrt{2}\ a^2&=f^1_1+f^1_3+h^1_1+h^1_3\ \ \ ,\ \ \ \sqrt{2}\ a^3=-f^2_0+f^2_2+h^2_0-h^2_2\ ,\cr
\sqrt{2}\ b_0&=f^2_0-f^2_2+h^2_0-h^2_2\ \ \ ,\ \ \ \sqrt{2}\ b_1=-f^1_0+f^1_2-h^1_0+h^1_2\ ,\cr
\sqrt{2}\ b_2&=-f^1_0+f^1_2+h^1_0-h^1_2\ \ \ ,\ \ \ \sqrt{2}\ b_3=-f^2_1-f^2_3-h^2_1-h^2_3\ ,\cr
\sqrt{2}\ c^0&=f^2_0+f^2_2-h^2_0-h^2_2\ \ \ ,\ \ \ \sqrt{2}\ c^1=-f^1_0-f^1_2+h^1_0+h^1_2\ ,\cr
\sqrt{2}\ c^2&=-f^1_0-f^1_2-h^1_0-h^1_2\ \ \ ,\ \ \ \sqrt{2}\ c^3=-f^2_1+f^2_3+h^2_1-h^2_3\ ,\cr
\sqrt{2}\ d_0&=f^2_1-f^2_3+h^2_1-h^2_3\ \ \ ,\ \ \ \sqrt{2}\ d_1=-f^1_1+f^1_3-h^1_1+h^1_3\ ,\cr
\sqrt{2}\ d_2&=-f^1_1+f^1_3+h^1_1-h^1_3\ \ \ ,\ \ \ \sqrt{2}\ d_3=f^2_0+f^2_2+h^2_0+h^2_2\ .}}
This gives an one--to--one map\foot{This is the restriction of the
map in \eqqs\resc\kds\ to the orbifold. Alternatively, the inverse relations
are:
\eqn\matchi{\eqalign{
f^1_0&=-\fc{\sqrt{2}}{4}\ (b_1 + b_2 + c^1 + c^2)\ \ \ ,\ \ \  
    f^1_1=\fc{\sqrt{2}}{4}\ (a^1 + a^2 - d_1 - d_2)\ \ \ ,\ \ \  
    f^1_2= \fc{\sqrt{2}}{4}\ (b_1 + b_2 - c^1 - c^2)\ ,\cr  
    f^1_3&= \fc{\sqrt{2}}{4}\ (a^1 + a^2 + d_1 + d_2)\ \ \ ,\ \ \  
    f^2_0= \fc{\sqrt{2}}{4}\ (-a^3 + b_0 + c^0 + d_3) \ \ \ ,\ \ \  
    f^2_1= -\fc{\sqrt{2}}{4}\ (a^0 + b_3 + c^3 - d_0)\ ,\cr 
    f^2_2&= \fc{\sqrt{2}}{4}\ (a^3 - b_0 + c^0 + d_3) \ \ \ ,\ \ \  
    f^2_3= -\fc{\sqrt{2}}{4}\ (a^0 + b_3 - c^3 + d_0) \ \ \ ,\ \ \  
    h^1_0= \fc{\sqrt{2}}{4}\ (-b_1 + b_2 + c^1 - c^2)\ ,\cr 
    h^1_1&= \fc{\sqrt{2}}{4}\ (-a^1 + a^2 - d_1 + d_2)\ \ \ ,\ \ \  
    h^1_2= \fc{\sqrt{2}}{4}\ (b_1 - b_2 + c^1 - c^2) \ \ \ ,\ \ \  
    h^1_3= \fc{\sqrt{2}}{4}\ (-a^1 + a^2 + d_1 - d_2)\ ,\cr 
    h^2_0&= \fc{\sqrt{2}}{4}\ (a^3 + b_0 - c^0 + d_3)\ \ \ ,\ \ \   
    h^2_1= \fc{\sqrt{2}}{4}\ (a^0 - b_3 + c^3 + d_0)\ \ \ ,\ \ \ 
    h^2_2= -\fc{\sqrt{2}}{4}\  (a^3 + b_0 + c^0 - d_3) \ ,\cr  
    h^2_3&= -\fc{\sqrt{2}}{4}\  (-a^0 + b_3 + c^3 + d_0)\ .}}} 
of the sixteen flux--components $a^i,b_i,c^i,d_i$ appearing
in the expansion \G\ and the sixteen gaugings $f_\Lambda^m,\ h^a_\Lambda\ ,\ \Lambda=0,\ldots,3\ ,\ 
a,m=1,2$.

Note, that in addition, there could be a dependence of the closed string superpotential 
on the modulus $T$ accounting for the $K3$--volume. 
The latter originates from the sections $F_\Lambda$.

Let us now discuss the dependence of the superpotential on the open string moduli $C^{7,a}$.
The quadratic piece $C^{7,a}$ in the sections $X^0,X^2$ appears as a matter of the symmetries
of the N=2 prepotential \prep. Eventually, as we shall see in the next section, these 
terms provide supersymmetric mass terms for the $D7$--brane fields $C^{7,a}$ describing
the transverse position along the torus $T^{2,3}$.
Therefore, let us move on to that part in \super, which stems from the sum 
$\Lambda=4,\ldots, 4+n_7$ or $a=1,\ldots,n_7$:
\eqn\partW{\eqalign{
W_2(C^{7,a},U^1,U^2)&=\sum_{a=1}^{n_7} C^{7,a}\ \lf\{\ f^1_a\ (U^1+U^2)
-f^2_a\ (1-U^1\ U^2)\ri.\cr
&\lf.\hskip2cm+h^1_a\ (U^1-U^2)-h^2_a\ (1+U^1 U^2)\ \ri\}\ .}}
As discussed around 
\eqq \kds, these gaugings are related to world--volume $2$--form fluxes 
on the $D7$--brane. The latter is wrapped around the $T^4/\IZ_2$ orbifold. Hence, let us briefly
introduce the $2$--form cohomology of this orbifold.

\def\jof{j_0}\def\omof{\om_0}
The toroidal orbifold $T^4/\IZ_2$ is the simplest of the orbifold limits of $K3$. 
The $\IZ_2$--action $(z_1,z_2)\to(-z_1,-z_2)$ has 16 fixed points, 
namely 
$(0,0),\,(0,\half),\,(0,\half i),\,(0,\half+\half i),\,(\half,0),\ldots,\,
(\half+\half i, \half+\half i)$. We know to have $h^{(2,0)}+h^{(1,1)}+h^{(0,2)}=1+20+1$ 
two-forms on $T^4/\IZ_2$.
The first six of them are the following:
\eqn\twoforms{\eqalign{
&dz^1\wedge dz^2,\quad d\ov z^1\wedge d\ov z^2,\cr
&dz^1\wedge d\ov z^2,\quad d\ov z^1\wedge d z^2,\cr
&dz^1\wedge d\ov z^1,\quad dz^2\wedge d\ov z^2\ .}}
The remaining 16 $(1,1)$--forms correspond to the collapsed $2$--cycles at the orbifold fixed 
points. We choose the complex structure 
\eqn\CS{
\om=dz^1\wedge dz^2\ ,}
which may be decomposed w.r.t to the real cohomology\foot{Instead, we may also choose the basis
\eqn\altbasis{\eqalign{
e_1&=\alpha_0-\beta^0\ ,\ e_2=\alpha_1-\beta^1\ ,\ e_3=\alpha_2-\beta^2\ ,\cr
e_4&=\alpha_0+\beta^0\ ,\ e_5=-\alpha_1-\beta^1\ ,\ e_6=\alpha_2+\beta^2\ ,}}
which shares the intersection properties of $SO(3,3)$, \ie $(e_i,e_i)=-2\ ,\ i=1,2,3$\ ,
$(e_i,e_i)=2\ ,\ i=4,5,6$, and $(e_i,e_j)=0\ ,i\neq j$.}
$(z^j=x^j+U^j y^j\ \ ,\ \ j=1,2)$
\eqn\realcoh{\eqalign{
\alpha_0=dx^1\wedge dx^2\ \ \ &,\ \ \ \beta^0=dy^1\wedge dy^2\ ,\cr
\alpha_1=dy^1\wedge dx^2\ \ \ &,\ \ \ \beta^1=-dx^1\wedge dy^2\ ,\cr
\alpha_2=dx^1\wedge dy^1\ \ \ &,\ \ \ \beta^2=-dx^2\wedge dy^2\ ,}}
with intersection matrix $M_i^{\ j}=\int\limits_{T^4/Z_2} \alpha_i\wedge\beta^j=\delta^j_i$, 
according to 
\eqn\decCS{
\omof=\alpha_0+\alpha_1\ U^1-\beta^1\ U^2+\beta^0\ U^1\ U^2\ =\sum_I 
\Pi^I \eta_I|_{T^2\times T^2/\ZZ_2}.}
As indicated, this is  the expansion of $\om$ in terms of the 
periods $\Pi^I$ in the notation of \defkt,
restricted to the subset $SO(2,2)\subset SO(3,19)$.
The K\"ahler form is chosen to be:
\eqn\Kaehler{
\jof=\sum_{j=1}^2 \im\Tc^j\  dx^j\wedge dy^j\ ,}
which is again a restriction  of the general expansion $j=\Pi^{I,3}\eta_I$,
to a 2 real dimensional subspace. 
The moduli ${\cal T}^j$ are the K\"ahler moduli in the string basis, see \eqq \relm.
The most general $2$--form flux $\Fc$ 
on the $D7$--brane world--volume, which is wrapped on $T^4/\IZ_2$, has 
six non--vanishing components $F^a_{ij}$ referring to the integral cohomology $H^2(T^4,\IZ)$ elements:
\eqn\mostgen{\eqalign{
\Fc^a&=\sum_{ij} \Fc^a_{ij}=\Fc^a_{NP}+\Fc^a_P\ ,\cr
\Fc^a_{NP}&=2\pi\ap\ \lf(\ F^a_{12}\ \alpha_2-F^a_{34}\ \beta^2\ \ri)\ ,\cr
\Fc^a_P&=2\pi\ap\ 
\lf(\ F^a_{13}\ \alpha_0+F^a_{23}\ \alpha_1-F^a_{14}\ \beta^1+F^a_{24}\ \beta^0\ri)\ .}}
We have split off a piece $\Fc^a_{NP}$ (\ie $\jof\wedge \Fc^a_{NP}\neq 0$) of the 
cohomology $H^{1,1}(T^4,\IZ)$, which is generically 
non-primitive in the restricted K\"ahler structure $\jof$ of \eqq\Kaehler. 
On the other hand, the part $\Fc^a_P$ is always primitive in the same K\"ahler structure (\ie
$\jof\wedge \Fc^a_{P}= 0$).
The flux components $F^a_{ij}$ obey the quantization rule: $F^a_{ij}=2\pi\ \fc{n^a_{ij}}{m^a_{ij}}$.

In the $F$--theory compactification on $X_4=\kb\times\kf$, the $2$--form flux $\Fc^a$ 
arises from the piece
\eqn\fourfl{
G_4=\sum_a  \Fc^a\wedge \etat_{a+4}\ ,}
with $\etat_{a+4}$ the $4\times4$ 2-forms on the exceptional spheres of the 4 $D_4$ singularities.
This flux induces the two potentials $W$ and $\Wh$ in \onceagain, associated with a 
four-dimensional $F$-term and $D$-term, respectively. On the $a$-th $D7$-brane, 
the flux integrals on the ``lower'' $K3$ become 
\eqn\SUPSS{\eqalign{
W|_{D7^a}:&\ 
\int_{\kb} \Fc_P^a \wedge \omof=
F_{24}^a-F_{14}^a\ U^1-F_{23}^a\ U^2+F_{13}^a\ U^1\ U^2\ ,\cr
d\Wh|_{D7^a}:&\ 
\int_{\kb} \Fc_{NP}^a 
\wedge \jof=\im \Tc^2\  F^a_{12}+\im \Tc^1\ F^a_{34}\ .}}
Thus the ``non--primitive'' part $\Fc_{NP}$ of the $2$--form flux gives rise 
to a $D$--term potential 
for the moduli $\cx T^j$ in \Kaehler, which vanishes if the condition \susy\
is met and $\cx F^a_{NP}$ is actually primitive at this point in the moduli.
The primitive part $\Fc_P$ gives rise to an $F$--term superpotential $W$ for the moduli $U^1,U^2$.
Moreover, the latter precisely reproduces the piece \partW\ 
of the superpotential, derived from gauged supergravity.
Hence, we recover the dictionary \kds\ for the case of the orbifold\foot{Note, 
that the gaugings $f^m_\Lambda,h^a_\Lambda$
refer to the lattice $SO(3,3)$, \ie the basis $e_i$ given in the last footnote.
This is way we obtain linear combinations as relations, which reflect the change of
basis from $\alpha_i,\beta^j$ to $e_i$.}, namely the supergravity gaugings 
$f^m_a,h^m_a,$ correspond to the flux--components as:
\eqn\Match{\eqalign{
f_a^1&=-\h\ (F_{14}^a+F_{23}^a)\ \ \ ,\ \ \ h_a^1=-\h\ (F_{14}^a-F_{23}^a)\ ,\cr
f_a^2&=-\h\ (F_{24}^a-F_{13}^a)\ \ \ ,\ \ \ h_a^2=-\h\ (F_{24}^a+F_{13}^a)\ ,\cr
f_a^3&=\h\ (F_{12}^a+F_{34}^a)\ \ \ ,\ \ \ h_a^3=-\h\ (F_{12}^a-F_{34}^a)\ .\cr
}}
{From} \eqqs \SUPSS\ we see, that the flux components 
$F^a_{12}, F^a_{34}$ give rise to a $D$--term potential, which 
fixes some of the K\"ahler moduli (\cf \eqq \fixed). On the other hand,
the remaining components $F^a_{13},F^a_{14},F^a_{23},F^a_{24}$ correspond to
the gaugings $f_a^{1,2},h_a^{1,2}$ and enter the $F$--term superpotential.
Eventually, the latter fix some of the complex structure moduli, as it has been 
recently also found in {\it Ref.} \AM.
To conclude, we have found the precise mapping of supergravity gaugings $f_\Lambda^m,h^a_\Lambda$
and fluxes in \tb string theory for the $T^4/\IZ_2\times T^2$ orientifold, 
for $\Lambda=0,1,\ldots,4+n_7$.

{}From our discussion of $F$- and $D$-terms in sect.~3.1 it follows that 
this is not the full story. Since we are studying
minima of a potential on the moduli space, we can not simply restrict to the
particular subset \decCS\ and \Kaehler\ of moduli. Instead one has to study the full
potential for all K3 moduli, including the ten geometrical untwisted moduli as well
as twisted moduli. In particular the flux $\cx F^a_{NP}$ gives a contribution 
to the $F$-term potential $W$ {\it and} the $D$ term potential $d\Wh$ in these moduli and  
similarly the flux $\cx F^a_{P}$ contributes to both potentials. The restriction to the
subset parametrized by $U^i$ and $T^i$ is consistent, if the minimization of
the full potential allows to eliminate all these extra moduli from the potential.\foot{This
happens if the extra moduli are either fixed to constant values at the minimum, or some
of the moduli decouple at the minimum and remain true moduli.}
In this case the truncated potential \SUPSS\ can be interpreted as the result of the minimization of
the full potential in all moduli except $T^i,U^i$, and the subsequent minimization of 
the moduli $T^i,U^i$ can then be studied using the truncated potentials.

According to the classification of complex structure preserving 
fluxes in section~3.1, the 2-form flux $\cx F^a_{NP}$ fulfills
the above consistency criterion, but the ansatz for $\cx F^a_P$ does not. Thus 
one can not fix complex structure moduli with this flux in the orientifold/orbifold. 
As described in section~3.4,
there is a different set of 2-form fluxes which yield a potential that does have a minimum 
on the restricted orbifold moduli space parametrized by $T^i,U^i$ and therefore 
can be consistently incorporated in the orbifold model. These fluxes couple to the 
off-diagonal and twisted moduli of the orbifold. On the other hand, 
the only way to fix the complex structure moduli within the simple orientifold/orbifold model 
based on the coset $SO(2,2)\times SO(1,1)$ is to use the bulk 3-form fluxes. 
Fixing the complex structure with bulk 3-form flux, 
this leaves an 18 dimensional moduli space of K\"ahler moduli which can not
be consistently fixed within the orbifold model based on $SO(2,2)\times SO(1,1)$.

Due to the block--diagonal form of $e$ (\cf \eqq \find), independent on the choice of
gaugings $f^m_\Lambda, h^a_\Lambda$, besides the vectormultiplet moduli $S,T,U,C^7,C^3$,
only the $SO(2,2)$--part of $e$, \ie 
the complex structure moduli $U^1,U^2$ of $T^4/\IZ_2$  may enter the superpotential $W$ 
through the Killing potentials  $P^1,P^2$. Hence, quite generically we obtain
a superpotential $W=W(S,U^1,U^2,U^3,C^{7,a},C^{3,a})$ depending on the fields $S, U^j, C^{7,a},C^{3,a}$.
This superpotential $W$, given in \eqq \superi, gives rise to the $F$--term potential 
\eqn\Fpot{
V_F=e^{K_V+K_H}(\ g^{i\ov j}\ D_iW D_{\ov j} \ov W-3\ |W|^2\ )\ ,}
with the index $i$ running over both the vectormultiplet moduli $S,T,U,C^7,C^3$
and the hypermultiplet moduli $U^1,U^2,T^1,T^2$.
Since the superpotential does not depend on the K\"ahler moduli, the scalar potential $V_F$ does
not depend on them either as a result of no--scale structure.
In addition, there is  a $D$--term scalar potential $V_D$, derived 
from the Killing potential $P^3$, given in \eqq \findkilling  
\def\srho{\sqrt{\rho}}
\eqn\Dpot{\eqalign{
V_D=g_{D_7}^2\cdot |\ P^3\ |^2&\sim\lf|\h\lf(\fc{1}{\srho}+\srho\ri)\ f^3_\Lambda+
\h\lf(\fc{1}{\srho}-\srho\ri)\ h^3_\Lambda\ri|^2 \ ,\cr
&\sim |\ (f^3_\La+h^3_\La)\, \im \Tc^1+(f^3_\La-h^3_\La)\, \im \Tc^2\ |^2\ ,
\phantom{\pmatrix{\big(\cr 1}}}}
in agreement with \eqq\sdtof. The above equations depends on the two K\"ahler moduli
in \Kaehler. The term $\sim f^3_\La$ is the universal $D$-term 
$j_0\wedge j_0$ for the orbifold, which drives the theory to the boundary
of the moduli. On the other hand,  
the case $h^3_\La\neq 0$, $\La>3$, gauges a $U(1)$ 
symmetry of $D7$--brane vectors
and gives rise to a non--vanishing $D$--term potential of the type \dex, which
is minimized for $\rho\equiv \fc{T^1_2}{T^2_2}=1$. Comparing with \eqqs 
\fixed\ and \SUPSS, this gauging may be identified with turning on the world--volume 
$2$--fluxes $F^a_{12},F_{34}^a$ on $D7$--branes.

Note, as pointed out already in \doubref\ADFL\ADFT,  
the condition $e=0$,\ \ie $L=1$ generically fixes all hypermultiplet moduli, except 
the volume $e^\Phi=\sqrt{T_2^1T_2^2}$, to
\eqn\fixing{
\rho=1\ \ \ ,\ \ \ |U^1|^2=|U^2|^2=1\ .}
\newsec{The $\mu$-term and soft SUSY breaking terms in the \tb $K3\times T^2$ orientifold}

In this section we study the $D=4$ effective N=1 supergravity action 
with spontaneous supersymmetry breaking due to non--vanishing auxiliary $F$--term 
components of the moduli fields. In this framework we calculate the flux--induced 
soft supersymmetry breaking terms of the effective four--dimensional $D7$--brane gauge theory
in lines of \SOFTref.

Compared to the effective approach performed in {\it Refs.} \doubref\LRSi\LRSii, the main 
difference is the superpotential \super\ depending now also on 
the open string moduli $C^{7,a}$ and $C^{3,b}$ referring to the $D7$-- and $D3$--brane
positions along the torus $T^2$. 
In particular, there is a supersymmetric mass term for the $D7$--brane moduli $C^{7,a}$
in the superpotential. The latter has been argued to be present by considering $F$--theory
compactified on $K3\times K3$ in {\it Ref.} \GKTT\ or after 
dimensional reduction of the Born--Infeld action of a single $D7$--brane in {\it Ref.} \CIU.
The dependence of the mass $m$ of the $D7$--brane modulus
$C^7$ on the $(0,3)$ and $(1,2)$ flux--components has been found to be 
\doubref\LRSi\LRSii
\eqn\msoftstackb{
m^2\sim \fc{1}{Y}\ \fc{1}{(S-\ov S)\ (U^3-\ov U^3)}\ \left\{\ \lf|\int G_3\wedge 
\Omega\,\ri|^2+
\lf|\int G_3\wedge \om_{B_1}\,\ri|^2+\lf|\int G_3\wedge \om_{B_2}\,\ri|^2\ \right\}\ ,}
with:
\eqn\Y{
Y=(S-\ov S)\ \prod_{j=1}^3 (T^j-\ov T^j)(U^j-\ov U^j)\ .}
Due to the presence of the supersymmetric mass term, depending linearly on the
gaugings $f_0^m,h_0^a$ and $f_2^m,h_2^a$, we expect an additional correction 
to the $D7$--brane scalar mass. This correction represents  a supersymmetric mass term
and is important to give a mass also to the fermionic partners of those moduli.

\subsec{Calculating the soft terms for $D7$--branes}

We first consider the case with just $D7$--branes (no $D3$--branes), no 2--form fluxes and set 
the $f^m_\Lambda$ and the $h^a_\Lambda$ to zero for $\Lambda=4,\ldots$, 
such that we are only left with the gaugings
$f^m_\Lambda,\ h^a_\Lambda$ with $\Lambda=0,...,3$. 
The precise mapping of these gaugings to the $3$--form 
flux components has been given in the previous section.

This case fits nicely into the formalism of \SOFTref. 
The soft terms are obtained by Taylor expanding the scalar potential 
around $C^{7a}=\ov C^{7a}=0$, i.e. in the following the $C^{7a},\ \ov C^{7a}$ are 
assumed to be small.

First, we will introduce some notation in order to make the results look less cluttered.
We want to cast our K\"ahler- and superpotential into the following form:
\eqn\KS{\eqalign{
K&=K_V+K_H=\hat K+ G_{C^{7a}\ov C^{7a}}\ C^{7a}\ov C^{7a}+(\h\ H_{C^{7a}\ C^{7a}}C^{7a}C^{7a}+
{\rm h.c.})+\ldots\ ,\cr
W&=\hat W+{1\over 2}\ \tilde\mu_{C^{7a} C^{7a}}\ C^{7a}C^{7a}+{1\over 3}\ \tilde Y_{ABC}\ C^{A}\ 
C^{B}C^{C}+\ldots\ .}}
Here, $\hat K$ and $\hat W$ refer to the respective potentials in the bulk and 
have the following form:
\eqn\hatKS{\eqalign{
\hat K&=-\sum_{M\in\{S, T^i,U^i\}} \ln(M-\ov M)=-\ln Y\ ,\cr
\hat W&={1\over \sqrt 2}\left\{(1-SU)A_0-(S+U)A_1-(1+SU)A_2-(U-S)A_3\right\}\ .}}
Here we introduced the shorthand
\eqn\short{
A_i=f^1_i(U^1+U^2)-f_i^2(1-U^1U^2)+h_i^1(U^1-U^2)-h_i^2(1+U^1U^2).}
The two functions $\hat K$ and $\hat W$ depend on the closed string moduli fields, only.
To be able to write $K=K_V+K_H$ in the form \KS, we expand the logarithm for small $C$. We then get:
\eqn\GH{
G_{C^{7a}\ov C^{7a}}={-1\over (S-\ov S)(U-\ov U)},\quad H_{C^{7a} C^{7a}}=
{1\over (S-\ov S)(U-\ov U)}=-G_{C^{7a}\ov C^{7a}}\ .}
The expression for $\tilde\mu_{C^{7a} C^{7a}}$ can be read off directly from the superpotential 
\super:
\eqn\tmu{
\tilde\mu_{C^{7a} C^{7a}}={1\over \sqrt 2}[A_0+A_2]\ .}
The remaining part of $W$ is zero in the case of only $D7$--branes and no $D3$--branes.

The scalar potential $V$ may be determined with the formula \Fpot:
\eqn\scalar{
V=e^K\ \lf[\ G^{I\ov J} (D_I W)(D_{\ov J} \ov W)-3\ |W|^2\ \ri]\ ,}
with the K\"ahler covariant derivative $D_I W=\p_I W+W\ \p_I K$ and the sum
running over all moduli $S,T^j,U^j,C^{7,a}$. 
The expansion of \scalar\ w.r.t. the matter field modulus $\phi\equiv C$ up to third order
may be arranged according to (\cf {\it Refs.} \SOFTref):
\eqn\arranged{\eqalign{
V&=\hat V(M,\ov M)+\h D^2+G^{C^{7a}\ov C^{7a}}\ \fc{\p W^{eff}}{\p{C^{7a}}}\ 
\fc{\p\ov W^{eff}}{\p{\ov C^{7a}}}\cr 
&+m_{C^{7a}\ov C^{7a}}^2\ |C^{7a}|^2+{1\over 3}\ A_{ABC}\ C^AC^BC^C
+\h\ B_{C^{7a}C^{7a}}\ (C^{7a})^2+\ldots\ .}}
The second and third terms of the first line preserve supersymmetry and
$W^{eff}$ may be considered as an effective N=1 superpotential, which
may take the following form:
\eqn\effW{
W^{eff}=\h\ \mu_{C^{7a}C^{7a}}\ (C^{7a})^2+\fc{1}{3}\ Y_{ABC}\ C^AC^BC^C+\ldots\ .}
Note that, as we shall see later, the quantities $\mu$ and $Y$ are not holomorphic.
Note especially that in the case of only $D7$--branes, all being transverse to the torus $T^{2,3}$,
the three--point functions between the matter fields $C^{7,a}$ are zero, \ie 
$0=\tilde Y_{ABC}=e^{-\hat K/2}Y_{ABC}$. 
{From} now on, we will for the sake of clearer notation suppress the superscript $a$ 
labeling the brane.

The first thing we need to calculate on the road to the soft terms are the $F$--terms. They 
can be written down in a very compact way as follows:
\eqn\Fterms{
\ov F^{\ov N}= e^{\hat K/2}\ \hat K^{\ov N L}\ \lf(\ \p_L\hat W+\hat W\p_L\hat K\ \ri)=Y^{-1/2}\ 
(N-\ov N)^2\ D_N\hat W\ .}
As an illustrative example, we display the expression for $\ov F^{\ov S}\, $:
$$\ov F^{\ov S}=Y^{-1/2}\ (S-\ov S){1\over \sqrt 2}
\left\{(1-\ov SU)A_0-(\ov S+U)A_1-(1+\ov SU)A_2-(U-\ov S)A_3\right\}.$$
Because of the exact correspondence between the superpotential from the gaugings and 
the flux superpotential $\int G_3\wedge \Omega$, we can express everything through flux integrals:
\eqn\fluxint{\hat W=\int G_3\wedge \Omega,\quad D_S \hat W\propto 
\int \ov G_3\wedge \Omega,\quad D_{U^i}\hat W\propto\int G_3\wedge \omega_{A_i}.}
Together with
\eqn\gravitino{
m_{3/2}=Y^{-1/2}\ \hat W}
we can now write down the expression for the bulk scalar potential.
In the no--scale case, the cosmological constant $\hat V(M,\ov M)$  consists of the 
$F$--terms of  the dilaton and complex structure moduli, only:
\eqn\cosmo{\eqalign{
\hat V(M,\ov M)&=\hat K_{S\ov S}\ F^S\ov F^{\ov S}+\sum_{j=1}^3
\hat K_{U^j\ov U^j}\ F^{U^j}\ov F^{\ov U^j}\cr
&=|Y|^{-1}\left\{\ \lf|\int \ov G_3\wedge \Omega\ri|^2+\sum_{i=1}^3\lf|\int G_3\wedge \omega_{A_i}
\ri|^2\ \right\}\ .}}
Before we can write down the scalar mass term, we must think about the curvature tensor, 
which is very simple in this case:
\eqn\rie{R_{S\ov S 3\ov 3}={-1\over (S-\ov S)^2}G_{C^{7}\ov C^{7}},\quad R_{U\ov U a\ov a}={-1\over (U-\ov U)^2}G_{C^{7}\ov C^{7}}.
}
All other components are zero. So the scalar masses become
\eqn\scalarmass{\eqalign{
m^2_{C^{7}\ov C^{7},\,{\rm soft}}&=\lf(\ |m_{3/2}|^2+\hat V\ \ri)\ 
G_{C^{7}\ov C^{7}}-\ov F^{M}\ F^{\ov N}\ R_{M\ov N 3 \ov 3}\cr
&=|Y|^{-1}\ \lf\{\ \lf|\!\int G_3\wedge \Omega\,\ri|^2+\lf|\!\int G_3\wedge \omega_{A_1}\ri|^2+
\lf|\!\int G_3\wedge \omega_{A_2}\ri|^2\ \ri\}\ G_{C^{7}\ov C^{7}}\ ,}}
which agrees with the form given in \msoftstackb.

Let us now determine the supersymmetric mass term $\mu_{C^{7}C^{7}}$, which gives
the mass $m_{C^{7}C^{7}}^f=\p^2_{C^{7}}\ W^{eff}=\mu_{C^{7}C^{7}}$ to the fermionic partners of the $D7$--brane
moduli $C^{7}$.
The $\mu$--term has the following form:
\eqn\softmu{\eqalign{
\mu_{C^{7}C^{7}}&=e^{\hat K/2}\tilde\mu_{C^{7}C^{7}}+m_{3/2}H_{C^{7}C^{7}}-\ov F^{\ov M}\ov \p_{\ov M}H_{C^{7}C^{7}}\cr
&=-{Y^{-1/2}\over \sqrt 2}\ G_{C^{7}\ov C^{7}}\ 
\left[(1-\ov S\ov U)A_0-(\ov S+\ov U)A_1-(1+\ov S\ov U)A_2-(\ov U-\ov S)A_3\right]\ .}}
In lines of the Giudice--Masiero effect \GM, in the above equation the $\tilde\mu$--term of \KS\ 
combines with supersymmetry breaking terms into an effective $\mu$--term 
in \effW. The latter gives a mass to the non--chiral fermions of the $D7$--brane moduli 
superfields. 

Naturally, we are now interested in learning which of the flux components contribute to $\mu$. 
We find that
\eqn\softmuflux{\eqalign{
\mu_{C^{7}C^{7}}&=-Y^{-1/2}\ {G_{C^{7}\ov C^{7}}}\ \int \ov G_3\wedge \omega_{A_3}\ .}}
This means that  only a $(2,1)$--flux component (which preserves supersymmetry)  contributes to $\mu$--term.


{From} the expansion of \scalar\ we find an additional (supersymmetric) 
contribution to $m_{C^{7}\ov C^{7}}^2$. This mass comes from 
$G^{C^{7}\ov C^{7}}\ \fc{\p W^{eff}}{\p{C^{7}}}\ \fc{\p\ov W^{eff}}{\p{\ov C^{7}}}$,  
after extracting
the order $|C^{7}|^2$, 
\eqn\key{\eqalign{
m^2_{C^{7}\ov C^{7},\,{\rm Susy}}&=\fc{i}{Y (S-\ov S)\ (U^3-\ov U^3)}\cr 
&\times \lf|(S-\ov S)\ (U^3-\ov U^3)\ \tilde\mu_{C^{7}C^{7}}-\hat W+
(S-\ov S)\ \fc{\p \hat W}{\p S}+(U^3-\ov U^3)\ \fc{\p \hat W}{\p U^3}\ri|^2\cr
&=G^{C^{7}\ov C^{7}}\ |\mu_{C^{7}C^{7}}|^2.
}}
The total mass term therefore amounts to:
\eqn\total{\eqalign{
m^2_{C^{7}\ov C^{7},\,{\rm soft+Susy}}&={G_{C^{7}\ov C^{7}}\over |Y|}
\left\{\lf|\int G_3\wedge \Omega\,\ri|^2+\lf|\int G_3\wedge \omega_{A_1}\ri|^2+
\lf|\int G_3\wedge\omega_{A_2}\ri|^2+\lf|\int \ov G_3\wedge \omega_{A_3}\ri|^2\right\}.}}

The trilinear couplings $A_{ABC}$ are zero in our case.
The $B$--term is obtained via the expansion of the scalar potential as seen in \arranged.
We find for $B$:
\eqn\Ba{\eqalign{
\half B_{C^{7}C^{7}}=&|Y|^{-1}\ \left\{H_{C^{7} C^{7}}
\left\{-\sum_{i=1}^3|\p_{U^i}\hat W(U^i-\ov U^i)|^2-|\p_S\hat W(S-\ov S)|^2\right.\right.\cr
&-4|\hat W|^2+(S-\ov S)(U-\ov U)
\,[\,\p_S\hat W\p_{\ov U}\hat{\ov W}+\p_U\hat W\p_{\ov S}\hat{\ov W}]\cr
&\left.+\left[\hat{\ov W}\,[\ \sum_{i=1}^3\p_{U^i}\hat W(U^i-\ov U^i)+
\p_s\hat W(S-\ov S)]+{\rm h.c.}\right]\right\}-2\tilde\mu\hat{\ov W}\cr
&-\tilde\mu\,[\,\p_{\ov U^1}\hat{\ov W}(U^1-\ov U^1)+\p_{\ov U^2}
\hat{\ov W}(U^2-\ov U^2)-\p_{\ov U}\hat{\ov W}(U-\ov U)-\p_{\ov S}\hat{\ov W}(S-\ov S)]\cr
&+\p_{U^1}\tilde\mu\p_{\ov U^1}\hat{\ov W}(U^1-\ov U^1)^2+\p_{U^2}\tilde\mu\p_{\ov U^2}\hat{\ov W}(U^2-\ov U^2)^2\cr
&\left.+\hat{\ov W}\,[\ \p_{U^1}\tilde\mu(U^1-\ov U^1)+\p_{U^2}\tilde\mu(U^2-\ov U^2)\,]\right\}.
}}
 In order to express the result in terms of flux integrals, we must find the flux expressions for $\tilde\mu$ and the $\p_{U^i}\tilde\mu$.
It is important to realize that $\tilde\mu=-\p_U\p_S\hat W.$ Writing $\hat W=\int G_3\wedge \Omega$ with $G_3=F_3-SH_3$, we find after close examination
\eqn\muflu{\eqalign{
(S-\ov S)(U-\ov U)\tilde\mu&=-(S-\ov S)\int \p_S G_3\wedge(U-\ov U) \p_U\Omega\cr
&=(S-\ov S)\int H_3\wedge (\Omega - \omega_{A_3})\cr
&=\int (\ov G_3- G_3)\wedge (\Omega - \omega_{A_3}).
}}
So the $NS$--part of the flux is picked out by the $S$--derivative and the $(0,3)$, $(3,0)$ and one of the $(1,2)$ and $(2,1)$--components are found to contribute to $\tilde\mu$.

$(S-\ov S)(U-\ov U)(U^i-\ov U^i)\p_{U^i}\tilde\mu,\ i=1,2$  needs to be expressed in terms of flux integrals, as well. By similar reasoning as above, we find the following flux integrals:
\eqn\dmu{\eqalign{
(S-\ov S)(U-\ov U)(U^1-\ov U^1)\p_{U^1}\tilde\mu&=\int(\ov G_3- G_3)\wedge (\Omega-\omega_{A_3}-\omega_{A_1}+\omega_{B_2}),\cr
(S-\ov S)(U-\ov U)(U^2-\ov U^2)\p_{U^2}\tilde\mu&=\int (\ov G_3-G_3)\wedge (\Omega-\omega_{A_3}-\omega_{A_2}+\omega_{B_1}).
}}
Together with \fluxint\ we have now collected everything necessary to write \Ba\ 
in terms of flux-integrals:
\eqn\Bflux{\eqalign{
\half B_{C^{7}C^{7}}&=|Y|^{-1}\ H_{C^{7} C^{7}}
\left\{-\sum_{j=1}^3\int \ov G_3 \wedge \omega_{A_i}\int \ov G_3\wedge \ov\omega_{A_i}-\int\ov G_3\wedge\Omega\int\ov G_3\wedge \ov\Omega\right.\cr
&+\int (\ov G_3-G_3)\wedge \omega_{A_3}\int \ov G_3\wedge \ov\Omega+\int (\ov G_3-G_3)\wedge \ov\Omega\int \ov G_3\wedge \omega_{A_3}\cr
&\left.+\int (\ov G_3-G_3)\wedge \ov\omega_{A_2}\int \ov G_3\wedge \ov\omega_{A_1}+\int (\ov G_3-G_3)\wedge\ov \omega_{A_1}\int \ov G_3\wedge \ov\omega_{A_2}\right\}.
}}


\subsec{Calculating the mass terms for $D7$--branes with 2--form fluxes}

We are interested exclusively in the 2--form fluxes corresponding to the angles of 
intersecting branes in the $T$--dual picture. 

When the 2--form fluxes, discussed in the previous section are turned on, 
the K\"ahler potential and therefore also the metric and $H$--term 
receive a flux dependence. The metric  and $H$--term for the $D7$--brane position modulus $C_3^7$
take the following form:

\eqn\twistmetric{G_{C^{7}_3\ov C^{7}_3}={-1\over{(U-\ov U)\ (S-\ov{S})}}+{\alpha'^{-2}f_{12} f_{34}\over(T-\ov T)(U-\ov U)},\quad H=-G.}
With this, we get the following curvature tensor components:
\eqn\fluxR{\eqalign{
R_{U\ov U3\ov 3}&={-1\over(U-\ov U)^2}\ G_{C^{7}_3\ov C^{7}_3},\cr
R_{S\ov S3\ov 3}&={2\alpha'^{-2}f_{12}f_{34}(S-\ov S)-(T-\ov T)\over(S-\ov S)^3(U-\ov U)[\alpha'^{-2}f_{12}f_{34}(S-\ov S)-(T-\ov T)]},\cr
R_{T\ov T3\ov 3}&={\alpha'^{-2}f_{12}f_{34}[2(T-\ov T)-\alpha'^{-2}f_{12}f_{34}(S-\ov S)]\over (T-\ov T)^3(U-\ov U)[\alpha'^{-2}f_{12}f_{34}(S-\ov S)-(T-\ov T)]},\cr
R_{T\ov S3\ov 3}&={-\alpha'^{-2}f_{12}f_{34}\over (S-\ov S)(T-\ov T)(U-\ov U)[\alpha'^{-2}f_{12}f_{34}(S-\ov S)-(T-\ov T)]}.
}}
This results in a changed soft scalar mass term:
\eqn\fluso{\eqalign{
m^2_{C^{7a}\ov C^{7a},\,{\rm soft}}=&|Y|^{-1}\ 
\left\{{-1\over(S-\ov S)(U-\ov U)}\left[\,\lf|\!\int G_3\wedge \Omega\,\ri|^2+
\lf|\!\int G_3\wedge \omega_{A_1}\,\ri|^2+\lf|\!\int G_3\wedge \omega_{A_2}\,\ri|^2
\right]\right.\cr
&+{\alpha'^{-2}f_{12}f_{34}\over(U-\ov U)[\alpha'^{-2}f_{12}f_{34}\ (S-\ov S)-(T-\ov T)]}
\left[\,\lf|\!\int G_3\wedge \Omega\,\ri|^2+\lf|\!\int G_3\wedge \omega_{A_1}\ri|^2\right.\cr
&\left.+\lf|\!\int G_3\wedge \omega_{A_2}\ri|^2+2\,\lf|\!\int \ov G_3\wedge \Omega\,\ri|^2
-\left(\int G_3\wedge \Omega\int G_3\wedge \ov\Omega+{\rm h.c.}\right)\right]\cr
&-{(\alpha'^{-2}f_{12}f_{34})^2\ (S-\ov S)\over(U-\ov U)(T-\ov T)[\alpha'^{-2}f_{12}f_{34}(S-\ov S)-(T-\ov T)]}\left[\,\lf|\!\int \ov G_3\wedge \Omega\,\ri|^2\right.\cr
&\left.\left.+\lf|\!\int G_3\wedge \omega_{A_1}\ri|^2+\lf|\!\int G_3\wedge \omega_{A_2}\ri|^2
\right]\right\}\ .}}
Turning on 2--form flux has therefore led to the contribution of more 3--form flux 
components to the mass terms than before, as had 
been realized already in \LRSi.
The  limit $f_{12},f_{34}\ra\infty$ changes the boundary conditions 
from Neumann to Dirichlet such that the $D7$--brane is converted into a $D3$--brane.
In this limit the piece quadratic in the two--form fluxes  is leading and describes
precisely the soft--breaking mass terms on the $D3$--brane \doubref\CIUold\GGJL.

The $\mu$--term is calculated as before, but with the modified $H$--term \twistmetric. 
This results in:
\eqn\muflu{\mu_{C^7C^7}=Y^{-1/2}\left[{-1\over(S-\ov S)(U-\ov U)}\int \ov G_3\wedge 
\omega_{A_3}+{\alpha'^{-2}f_{12}f_{34}
\over(T-\ov T)(U-\ov U)}\int G_3\wedge \omega_{A_3}\right].}
In the limit $f_{12},f_{34}\ra\infty$ we again recover the $D3$--brane result of only
$(1,2)$--form fluxes contributing to $\mu$ \doubref\CIUold\GGJL.
The contribution of the $\mu$--term to the total mass term is $G^{C^7\ov C^7}|\mu|^2$:
\eqn\flumusq{\eqalign{
G^{C^7\ov C^7}|\,\mu_{C^7C^7}|^2=&|Y|^{-1}\ 
\left\{{-1\over(S-\ov S)(U-\ov U)}\ \lf|\!\int \ov G_3\wedge \omega_{A_3}\,\ri|^2\right.\cr
&+{\alpha'^{-2}f_{12}f_{34}\over(U-\ov U)\ [-\alpha'^{-2}f_{12}f_{34}\ (S-\ov S)-(T-\ov T)]}
\left[-\lf|\!\int \ov G_3\wedge \omega_{A_3}\,\ri|^2\right.\cr
&\left.+\lf(\int\ov G_3\wedge\omega_{A_3}\int\ov G_3\wedge\ov\omega_{A_3}+{\rm h.c.}\ri)
\right]\cr
&\left.+{(\alpha'^{-2}f_{12}f_{34})^2(S-\ov S)\over(U-\ov U)(T-\ov T)
[\alpha'^{-2}f_{12}f_{34}(S-\ov S)-(T-\ov T)]}\,\lf|\!\int G_3\wedge 
\omega_{A_3}\,\ri|^2\right\}\ .}}
Notice that not the entire contribution of the $\mu$--term originates from $(2,1)$--form
fluxes anymore.  Only the terms
\eqn\suflu{\eqalign{& |Y|^{-1}
\left\{{-1\over(S-\ov S)(U-\ov U)}\ \lf|\!\int \ov G_3\wedge \omega_{A_3}\,\ri|^2\right.\cr
&\left.-{\alpha'^{-2}f_{12}f_{34}\over(U-\ov U)[-\alpha'^{-2}f_{12}f_{34}(S-\ov S)-(T-\ov T)]}\ 
\lf|\!\int \ov G_3\wedge \omega_{A_3}\,\ri|^2\right\}}}
proportional to $|\!\int \ov G_3\wedge \omega_{A_3}\,|^2$ preserve supersymmetry.
The total mass term is $m^2_{\rm total}=m^2_{soft}+G^{C^7\ov C^7}|\ \mu\ |^2$.

To conclude, we have derived the soft susy--breaking mass terms and an effective $\mu$--term, 
which depends on $(2,1)$--form fluxes 
only in the case of vanishing
world--volume $2$--form flux on the $D7$--branes. 
In the presence of world--volume $2$--form fluxes, the effective $\mu$--term receives a 
$(1,2)$--dependent piece.
On the other hand, the masses $m_{C^7\ov C^7}$ of 
scalar fields of $D7$--brane moduli, describing their positions,
receive both $3$-- and $2$--form flux dependent contributions.

\newsec{Conclusions}

In this paper we have discussed the superpotential and the vacuum
structure of F-theory on $K3\times K3$ with general 4-form
flux turn on. In string theory this corresponds to \tb on
$K3\times T^2$ with non-constant dilaton, with non-vanishing 3-form
flux in the closed (bulk) string sector and with $D7$--branes equipped
with non-zero open string 2-form flux.
As a result of our discussion it turns out that for generic fluxes
all open and closed string moduli are fixed, except for a 
volume factor.\foot{Note, that this statement includes also the twisted sector of
orbifold limits. Generalized flux compactifications, where
different types of moduli can be fixed to various degrees have been 
recently discussed  also in other setups, in particular for all
but a single volume  modulus in \AM, for all geometric bulk 
moduli in \doubref\KKP\GL, and for a subset of orbifold moduli in \DKPZ.}
Furthermore the minimization of the open string $D7$--brane
moduli is such that the orientifold is generically unstable, i.e.
that the positions of the $D7$--branes are different from
their orientifold limits (and that the dilaton is non-constant in
a generic F-theory flux vacuum). However we also showed that there exist
particular choices of F-theory fluxes which drive the theory to a stable
orientifold limit. We also compare these results with an effective  
supergravity 
formulation of the flux compactification in terms of ${\cal N}=2$
gauged supergravity.

This last observation is important for the second part of the paper
where we discuss the $\mu$-term and the softly supersymmetry breaking terms
on the $D7$--brane world-volume in the orientifold limit, and in particular
in the orbifold limit of $K3$.
The result of this discussion is that the effective supersymmetric
mass terms of non-chiral fermions of the $D7$--brane moduli fields
depend both on (2,1)-form fluxes as well as on (1,2)-form
fluxes in case of non-vanishing 2-form fluxes on the $D7$--branes.
In addition, 
we should emphasize that these results 
should be used in the following way:
the derived equations for the $\mu$-term and the soft terms are parametrized
by the values of the moduli fields ($S,T,U$) and also by the values
of the 2-form and 3-form fluxes. However in any vacuum the
moduli are generically
fixed and hence are functions of a given choice of 2-form and
3-form fluxes. Therefore, in order to derive
actual (numerical) values for the soft terms, one first has to determine
the vacuum in any given flux compactification, i.e. the fixed values
for moduli, and use them as input in the equations for the $\mu$-term
and the soft terms. Of course there might exist special combinations of
soft term (certain ratios etc) which are essentially moduli
independent and therefore do not require the full 
determination of the non-supersymetric  vacuum structure.

\vskip20pt

\noi{\bf Acknowledgements:} We are grateful to 
H. Jockers and J. Louis for valuable discussions and informing us about 
their related work prior to publication. We also thank 
M. Trigiante for valuable discussions. This work is supported in part by the
Deutsche Forschungsgemeinschaft as well as by the
EU-RTN network {\sl Constituents, Fundamental Forces and Symmetries
of the Universe} (MRTN-CT-2004-005104). S.R. 
thanks the university of Munich for hospitality.

\vskip20pt
\goodbreak

\appendix{A}{Some general facts on K3}
In the following we collect some facts on the cohomology $H^*(X)$ of
a K3 $X$ and period integrals defined on it; we refer to 
\Ak\Mor\ for more background material.
The integral cohomology $H^{even}(X)$ has the structure of a 
selfdual lattice $\Gamma^{4,20}$ of signature $-16$. Splitting off 
a factor $H^0\oplus H^4$ of signature (1,1), the orthogonal sub-lattice 
$\Gamma^{3,19}\subset \Gamma^{4,20}$ corresponds to 
the 2-form cohomology, generated by 3 selfdual and
19 anti-selfdual 2-forms. Fixing an embedding $H^2(X,\ZZ)\to \Gamma^{3,19}$
and choosing a basis  $\{\eta_I\}$  for $H^2(X,\ZZ)$,
the inner product on $\Gamma^{3,19}$ is defined by
\eqn\defip{
M_{IJ}=\int \eta_I\wedge \eta_J=
\pmatrix{U&&&&&\cr&U&&&&\cr&&U&&&\cr&&&-C_{E8}&&\cr&&&&-C_{E8}&\cr},
\ \ \ U=\pmatrix{0&1\cr 1&0},
}
where $-C_{E_8}$ denotes the negative of the Cartan matrix for $E_8$.
An Einstein metric at volume one is specified, 
up to discrete identifications,  by choosing a space-like 3-plane 
in $\IR^{3,19}\supset\Gamma^{3,19}$
spanned by the 3 orthogonal, selfdual forms 
$\om^i$, $i=1,..,3$. Taking into account discrete identifications
corresponding to isometries of $\Gamma^{3,19}$, 
the moduli space is
\eqn\eins{
O(\Gamma^{3,19})\backslash O(3,19)/ O(3)\times O(19)\times \IR_+,}
where the last factor parametrizes the volume.

Fixing a direction in the $SU(2)$ of 
complex structures, the $\om^i$ combine into a holomorphic 2-form,
say $\om=\om^1+i\om^2$ and the K\"ahler form $j=\om^3$ as in \triplet;
they fulfill
\eqn\tfcon{
\om\wedge \om=j\wedge \om=0, \qquad \int j\wedge j >0,\qquad \int \om\wedge \bb \om >0.
}
These 2-forms may be expanded as 
$
\om = \Pi^I\eta_I$, $j=\Pi^{3,I}\eta_I,
$
with expansion coefficients given by the period integrals
$$
\Pi^I=\int_{\gamma_I}\om = \int \om\wedge \eta^I,\qquad \Pi^{3,I}=\int_{\gamma_I}j,
$$
where $\{\ga_I\}$ denote the basis of $H_2(X,\ZZ)$ dual to the $\{\eta_I\}$ and
indices are raised and lowered with the help of the metric \defip\ and its inverse.

{}From the first equation in \tfcon, the periods 
$\Pi^I$ satisfy the quadratic constraint
\eqn\perqua{\Pi^I\, M_{IJ} \, \Pi^J=0.}
The independent periods may serve as local projective coordinates
on the moduli space of complex structures.

The Picard group of $X$ is defined as the group of integral (1,1) forms
\eqn\defpic{\Pic(X)=H^2(X,\ZZ)\cap H^{1,1}(X),}
which implies $\int \om\wedge \eta_I=0$ for $\eta_I \in \Pic(X)$.
For generic complex structure, $\Pic(X)$ is trivial. However for the elliptic 
fibered $\kf$ with section, the duals of the elliptic fiber $E$ and the 
dual of the section $B$ are elements of $\Pic(X)$ with intersections
\ebi. The complex structure is determined
by the period vector $\Pi^I$ on the remaining 20 basis elements.
For the following it is sometimes convenient to use an
orthogonal basis for the lattice, with intersection matrix
\eqn\defimo{
M_{IJ}=\rmx{diag}(2,2,-2,-2,-1^{16})\ .\phantom{\pmatrix{1\cr1}}
}
In this basis the period vector reads 

\eqn\pari{
\Pi_I = \pmatrix{1- S  U +\h C^aC^a\cr  S + U \cr 
1+ S  U -\h C^aC^a\cr- S + U \cr -C^a\cr}\ ,
}

\noi where we have fixed the projective action in the patch where 
$\Pi_1+\Pi_3\neq 0$ and we have also used the constraint \perqua.

The period vector \pari\ describes a set of standard coordinates on
the coset $SO(2,18)/(SO(2)\times SO(18))$. Note that in this
parametrization an $SO(2,2;\ZZ)$ of reparametrizations is realized
linearly on the periods $ S $ and $ U $. The same period vector
also appears as the upper half of a special choice for the 
symplectic section of $\cx N=2$ supergravity based on the special 
K\"ahler manifold \msugra. In particular it has been proposed in 
\ADFT\ that this section describes the special geometry for the 
type IIB orientifold on $T^2/\ZZ_2$. However despite of the similarity,
the period vector \pari\ at $C^a=0$ cannot be identified with
the symplectic section for the type IIB orientifold on $T^2/\ZZ_2$.
This follows e.g. from the uniqueness of the selfdual lattice
with signature $(2,18)$ which implies that the vanishing cycles at $C^a=0$
must span the root lattice of $E_8\times E_8$. The proper symplectic
section is described by a transformation from \pari\ to a different
basis that is described in app. B below.

\appendix{B}{Orientifold limit $\xof$ of $\kf$}
The non-abelian gauge symmetry of the orientifold limit of type IIB
on $T^2/\ZZ_2$ is $SO(8)^4$ from 4 D7-branes on top of each of 
the orientifold planes. To describe the corresponding 
singular K3 $\xof$, we
need a parametrization different from \pari, such that the 
vanishing cycles correspond to four copies of root lattice of $D_4$.
Note that the orientifold limit is also not described by the simple 
orbifold $T^2\times T^2/\ZZ_2$, whose vanishing cohomology  
has intersection matrix $A_1^{16}$, but by resolving
three $A_1$ singularities in a different orbifold denoted by 
$T^4/\hx D_4'$\foot{Strictly speaking
these comments refer to the manifold $\kf$ that appears 
in the M-theory compactification
to 7 dimensions, before taking the F-theory limit.}. The description of
the generators of the integral cohomology of K3 orbifolds in terms of the
vanishing cohomology and the cohomology inherited from $T^4$ is quite 
delicate. We refer to \KW\ for a complete discussion of
the cohomology of K3 orbifolds without translations as well as 
for references to the mathematical literature on the subject. 

In the following we
restrict ourselves to describe a parametrization of the period vector 
in an appropriate basis for the orientifold which does however not generate
$H^2(\xof,\ZZ)$ over the integers. The integrality properties may be recovered
from transforming integral fluxes in the basis corresponding to \pari\ to
the new basis described below.

To find the appropriate basis change, we may borrow from the 
results on the dual heterotic string compactified on $T^2$ with
Wilson lines \joe\GKTT. In the orthogonal basis with intersection matrix
\defimo, the roots of a single $E_8$ factor correspond to 
the 240 vectors of the form
\def\smallh{{\scriptstyle \fc{1}{2}}}
\eqn\latv{
(\underline{\pm 1,\pm 1,0,0,0,0,0,0}),\qquad 
(\pm\smallh,\pm\smallh,\pm\smallh,\pm\smallh,
\pm\smallh,\pm\smallh,\pm\smallh,\pm\smallh),
}
with all possible choices of signs for the l.h.s. and an even
number of minus sign for the r.h.s. These roots 
become orthogonal to $\om$ in the limit $C^a=0,$ $a=1,...,8$ in \pari.
Similarly a different copy of $E_8$ roots becomes orthogonal to
$\om$ if in addition $C^a=0,$ $a=9,...,16$. In the following
the two $E_8$ factors enter symmetrically.

To describe the limit with roots in $4\ D_4$ instead, we first change
the complex structure from $C^a=0$ to
\eqn\vshift{
v=(1,0,0,0,\fc{ U }{2},\fc{ U }{2},\fc{ U }{2},\fc{ U }{2}).
}
It is readily verified that the roots orthogonal to $\om$ span two factors
of the $D_4$ root lattice for any $ U $. 
Including the second $E_8$ factor and 
after change of basis and shifting the coordinates
one arrives at a new ``period vector''
\eqn\parii{\displaystyle{
\Pi_I=\pmatrix{
1- S  U +\h C^aC^a\phantom{\pmatrix{1\cr2}\hskip-20pt}\cr 
 S + U \phantom{\pmatrix{1\cr2}\hskip-20pt}\cr
-1- S  U -\h C^aC^a\phantom{\pmatrix{1\cr2}\hskip-20pt}\cr
- S + U \phantom{\pmatrix{1\cr2}\hskip-20pt}\cr
-C^a}\ .
}}
Let us stress that 
although the period vector looks
formally identical to \pari, it is not so for two reasons. Firstly the
locus $C^a=0$ corresponds to a vanishing cohomology 
given by the root lattice $4\ D_4$.
Secondly the basis vectors do {\it not} span the integral lattice
over the integers. See \KW\ for an appropriate basis of generators
for the integral lattice in terms of exceptional divisors and invariant
forms. However one may also reconstruct the integral
vectors from the known lattice vectors in the basis \pari\ 
and the known transformation from \pari\ to \parii. One may also
verify explicitely that there is a choice of integral flux that 
drives the theory to the orientifold limit above, or also the 
orbifold $T^4/\ZZ_2$ discussed in the text.

\appendix{C}{Three--form fluxes of $T^4/\IZ_2\times T^2$}

In this appendix we present the (untwisted) three--form cohomology, used before to describe the 
three--form flux $G_3$ of the \tb orientifold compactification $T^4/\IZ_2\times T^2$.
For toroidal orbifolds $T^6/\IZ_N$ the complex structures are introduced through:
\eqn\achieved{
dz^j=\sum_{i=1}^3\ \rho^j_i\ dx^i+ \tau^j_i\ dy^i,\ \ \ j=1,2,3\ .}
For the case of the  direct product $T^6=T^4/\IZ_2\times T^2$ we may choose
$\rho^j_i=\delta^j_i$ and $\tau^3_i=\delta^3_i\ U^3$.
In addition, for the present case, the $T^4$ being itself  the direct product
of two two--tori $T^4=T^{2,1}\times T^{2,2}$, we have $\tau_i^j=\delta_i^j\ U^j$.
To this end, we have: $dz^i=dx^i+U^idy^i\ ,\ i=1,2,3$.

The 3--forms of $T^4/\IZ_2\times T^2$ are obtained by wedging the one--forms $dz^3$ or $d\ov z^3$ 
of the torus $T^{2,3}$ to the 2--forms given in \twoforms, 
which means that we end up with twelve  3--forms from
the untwisted sector:
\eqn\cplxz{\eqalign{
\om_{A_0}&=d\ov z^1\wedge d\ov z^2\wedge d\ov z^3\ \ \ ,\ \ \ \ 
\om_{B_0}=dz^1\wedge d z^2\wedge d z^3,\cr
\om_{A_1}&=d\ov z^1\wedge dz^2\wedge dz^3\ \ \ ,\ \ \ 
\om_{B_1}= dz^1\wedge d\ov z^2\wedge d\ov z^3, \cr 
\om_{A_2}&=dz^1\wedge d\ov z^2\wedge dz^3\ \ \  ,\ \ \ \  
\om_{B_2}=d\ov z^1\wedge dz^2\wedge d\ov z^3,\cr
\om_{A_3}&=dz^1\wedge dz^2\wedge d\ov z^3\ \ \  ,\ \ \ 
\om_{B_3}= d\ov z^1\wedge d\ov z^2\wedge d z^3,\cr
\om_{A_4}&=dz^1\wedge d\ov z^1\wedge dz^3\ \ \ ,\ \ \ \  
\om_{B_4}=d \ov z^1\wedge dz^1\wedge d\ov z^3,\cr
\om_{A_5}&=dz^2\wedge d\ov z^2\wedge dz^3\ \ \ ,\ \ \ 
\om_{B_5}= d\ov z^2\wedge d z^2\wedge d\ov z^3\ .}}
The forms $\om_{A_0},\ldots,\om_{A_3}$ and $\om_{B_0},\ldots,\om_{B_3}$ fulfill the 
primitivity condition $J\wedge \om=0$, 
with the K\"ahler form $J=\sum\limits_{j=1}^3\im(\Tc^j)\ dz^j\wedge d\ov z^j$.
On the other hand, the forms $\om_{A_4}, \ \om_{A_5},\ \om_{B_4},\ \om_{B_5}$ are not primitive, \ie
$J\wedge \om\neq 0$, 
yet it is possible to form two primitive linear combinations from them:
\eqn\primitive{
\om_{A_p}=\om_{A_5}-{\im(\Tc^1)\over \im(\Tc^2)}\ \om_{A_4}\ ,\quad 
\om_{B_p}=\om_{B_5}-{\im(\Tc^1)\over \im(\Tc^2)}\ \om_{B_4}\ .}
In \eqq \G\ the flux $G_3$ is expanded w.r.t. to the real cohomology:
\eqn\realbase{
\eqalign{\alpha_0&=dx^1 \wedge dx^2  \wedge dx^3\ \ \ ,\ \ \ 
\beta^0=dy^1 \wedge dy^2 \wedge dy^3\ ,\cr
\alpha_1&=dy^1 \wedge dx^2  \wedge dx^3\ \ \ ,\ \ \ \beta^1=-dx^1 \wedge dy^2\wedge dy^3\ ,\cr
\alpha_2&=dx^1 \wedge dy^2  \wedge dx^3\ \ \ ,\ \ \ \beta^2=-dy^1 \wedge dx^2 \wedge dy^3\ ,\cr
\alpha_3&=dx^1 \wedge dx^2  \wedge dy^3\ \ \ ,\ \ \ \beta^3=-dy^1 \wedge dy^2 \wedge dx^3\ ,\cr
\alpha_4&=dx^1 \wedge dy^1  \wedge dx^3\ \ \ ,\ \ \ \beta^4=-dx^1 \wedge dy^1 \wedge dy^3\ ,\cr
\alpha_5&=dx^2 \wedge dy^2  \wedge dx^3\ \ \ ,\ \ \ \beta^5=-dy^2 \wedge dy^2 \wedge dy^3\ .}}
In this basis, the forms $\alpha_4,\alpha_5,\beta^4$ and $\beta^5$ are non--primitive.
At any rate, three--form flux components w.r.t. the latter 
do not give a contribution in the superpotential \TVW.
The $(3,0)$--form may be expanded w.r.t. the basis \realbase:
\eqn\expOmeg{\eqalign{
\Omega&=\alpha_0+U^1\ \alpha_1+U^2\ \alpha_2+U^3\ \alpha_3\cr
      &-U^2\ U^3\ \beta^1-U^1\ U^3\ \beta^2-U^1\ U^2\ \beta^3+U^1\ U^2\ U^3\ \beta^0\ .}}

\ninerm
\listrefs
\end